\documentclass[structabstract,longauth]{aa}  
\usepackage[stable]{footmisc}
\usepackage{graphicx}
\usepackage{color}
\usepackage{colortbl}
\usepackage{amssymb}
\usepackage{amsmath}
\usepackage{longtable}
\usepackage{txfonts}
\usepackage{multirow}
\usepackage{textcomp}
\usepackage{natbib}
\bibpunct{(}{)}{;}{a}{}{,} 
\usepackage{ulem}

\begin{document}

\title{The VIMOS Public Extragalactic Redshift Survey (VIPERS)\thanks{based on observations collected at the European Southern Observatory, Cerro Paranal, Chile, using the Very Large Telescope under programme 182.A-0886 and partly 070.A-9007. 
Also based on observations obtained with MegaPrime/MegaCam, a joint project of CFHT and CEA/DAPNIA, at the Canada-France-Hawaii Telescope (CFHT), which is operated by the National Research Council (NRC) of Canada, the Institut National des Sciences de l’Univers of the Centre National de la Recherche Scientifique (CNRS) of France, and the University of Hawaii. 
This work is based in part on data products produced at TERAPIX and the Canadian Astronomy Data Centre as part of the Canada-France-Hawaii Telescope Legacy Survey, a collaborative project of NRC and CNRS. 
The VIPERS web site is http://www.vipers.inaf.it/.}}
\subtitle{A~support vector machine classification of galaxies, stars, and~AGNs}
\titlerunning{VIPERS: A~SVM classification of galaxies, stars, and~AGNs}

\author{K.~Ma{\l}ek\thanks{Postdoctoral Fellow of the Japan Society for the Promotion of Science}\inst{1} \and
A.~Solarz\inst{1} \and
A.~Pollo\inst{2,3} \and 
A.~Fritz\inst{4} \and  
B.~Garilli\inst{4,5} \and   
M.~Scodeggio\inst{4} \and  
A.~Iovino\inst{6} \and   
B.~R.~Granett\inst{6} \and  
U.~Abbas\inst{7} \and  
C.~Adami\inst{5} \and       
S.~Arnouts\inst{8,5} \and  
J.~Bel\inst{9} \and  
M.~Bolzonella\inst{10} \and  
D.~Bottini\inst{4} \and  
E.~Branchini\inst{11,12,13} \and 
A.~Cappi\inst{10} \and  
J.~Coupon\inst{14} \and   
O.~Cucciati\inst{10} \and  
I.~Davidzon\inst{10,15} \and  
G.~De~Lucia\inst{16} \and  
S.~de~la~Torre\inst{17} \and  
P.~Franzetti\inst{4} \and  
M.~Fumana\inst{4} \and  
L.~Guzzo\inst{6,18} \and  
O.~Ilbert\inst{5} \and  
J.~Krywult\inst{19} \and  
V.~Le~Brun\inst{5} \and
O.~Le~Fevre\inst{5} \and  
D.~Maccagni\inst{4} \and  
F.~Marulli\inst{15,20,10} \and  
H.~J.~McCracken\inst{21} \and  
L.~Paioro\inst{4} \and  
M.~Polletta\inst{4} \and  
H.~Schlagenhaufer\inst{22,23} \and  
L.~A.~M.~Tasca\inst{5} \and  
R.~Tojeiro\inst{24} \and  
D.~Vergani\inst{25} \and  
A.~Zanichelli\inst{26} \and  
A.~Burden\inst{24} \and  
C.~Di~Porto\inst{10} \and  
A.~Marchetti\inst{27,6} \and  
C.~Marinoni\inst{9,28} \and  
Y.~Mellier\inst{21} \and  
L.~Moscardini\inst{15,20,10} \and  
R.~C.~Nichol\inst{24} \and  
J~.A.~Peacock\inst{17} \and 
W.~J.~Percival\inst{24} \and  
S.~Phleps\inst{23} \and  
M.~Wolk\inst{21} \and  
G.~Zamorani\inst{10}}      

\offprints{K.~Ma{\l}ek\\ \email{malek.kasia@nagoya-u.jp}}
\institute{
	Department of Particle and Astrophysical Science, Nagoya University, Furo-cho, Chikusa-ku, 464-8602 Nagoya, Japan 
	\and Astronomical Observatory of the Jagiellonian University, Orla 171, 30-001 Cracow, Poland 
	\and National Centre for Nuclear Research, ul. Hoza 69, 00-681 Warszawa, Poland 
	\and INAF - Istituto di Astrofisica Spaziale e Fisica Cosmica Milano, via Bassini 15, 20133 Milano, Italy 
	\and Aix Marseille Universit\'e, CNRS, LAM (Laboratoire d'Astrophysique de Marseille) UMR 7326, 13388, Marseille, France  
	\and INAF - Osservatorio Astronomico di Brera, Via Brera 28, 20122 Milano, via E. Bianchi 46, 23807 Merate, Italy 
	\and INAF - Osservatorio Astrofisico di Torino, 10025 Pino Torinese, Italy 
	\and Canada-France-Hawaii Telescope, 65--1238 Mamalahoa Highway, Kamuela, HI 96743, USA 
	\and Aix-Marseille Universit\'e, CNRS, CPT (Centre de Physique  Th\'eorique) UMR 7332, F-13288 Marseille, France 
	\and INAF - Osservatorio Astronomico di Bologna, via Ranzani 1, I-40127, Bologna, Italy 
	\and Dipartimento di Matematica e Fisica, Universit\`{a} degli Studi Roma Tre, via della Vasca Navale 84, 00146 Roma, Italy 
	\and INFN, Sezione di Roma Tre, via della Vasca Navale 84, I-00146 Roma, Italy 
	\and INAF - Osservatorio Astronomico di Roma, via Frascati 33, I-00040 Monte Porzio Catone (RM), Italy 
	\and Institute of Astronomy and Astrophysics, Academia Sinica, P.O. Box 23-141, Taipei 10617, Taiwan 
	\and Dipartimento di Fisica e Astronomia - Universit\`{a} di Bologna, viale Berti Pichat 6/2, I-40127 Bologna, Italy 
	\and INAF - Osservatorio Astronomico di Trieste, via G. B. Tiepolo 11, 34143 Trieste, Italy 
	\and SUPA, Institute for Astronomy, University of Edinburgh, Royal Observatory, Blackford Hill, Edinburgh EH9 3HJ, United Kingdom 
	\and Dipartimento di Fisica, Universit\`a di Milano-Bicocca, P.zza della Scienza 3, I-20126 Milano, Italy 
	\and Institute of Physics, Jan Kochanowski University, ul. Swietokrzyska 15, 25-406 Kielce, Poland 
	\and INFN, Sezione di Bologna, viale Berti Pichat 6/2, I-40127 Bologna, Italy 
	\and Institute d'Astrophysique de Paris, UMR7095 CNRS, Universit\'{e} Pierre et Marie Curie, 98 bis Boulevard Arago, 75014 Paris, France 
	\and Universit\"{a}tssternwarte M\"{u}nchen, Ludwig-Maximillians Universit\"{a}t, Scheinerstr. 1, D-81679 M\"{u}nchen, Germany 
	\and Max-Planck-Institut f\"{u}r Extraterrestrische Physik, D-84571 Garching b. M\"{u}nchen, Germany 
	\and Institute of Cosmology and Gravitation, Dennis Sciama Building, University of Portsmouth, Burnaby Road, Portsmouth, PO1 3FX, United Kingdom 
	\and INAF - Istituto di Astrofisica Spaziale e Fisica Cosmica Bologna, via Gobetti 101, I-40129 Bologna, Italy 
	\and INAF - Istituto di Radioastronomia, via Gobetti 101, I-40129, Bologna, Italy 
	\and Universit\`{a} degli Studi di Milano, via G. Celoria 16, 20130 Milano, Italy 
}

\authorrunning{K. Ma{\l}ek et al.,}
 
\date{Received 11 March 2013 /Accepted 26 June 2013}
 
\abstract
   {}
   {The aim of this work is to develop a comprehensive method for classifying sources in large sky surveys and to apply the techniques to the VIMOS Public Extragalactic Redshift Survey (VIPERS). 
Using the optical (u$^*$, g', r', i') and NIR data (z',  K$_{s}$), we develop a classifier, based on broad-band photometry, for identifying stars, AGNs, and galaxies, thereby improving the purity of the VIPERS sample.
}
   {
Support vector machine (SVM) supervised learning algorithms  allow the automatic classification of objects into two or more classes based on a multidimensional parameter space.  
In this work, we tailored the SVM to classifying stars, AGNs, and galaxies and applied this classification to the VIPERS data.
We trained the SVM using  spectroscopically confirmed sources from the VIPERS and VVDS surveys.
}
{ 
We tested two SVM classifiers and conclude that including NIR data can significantly improve the efficiency of the classifier. 
The self-check of the best optical + NIR classifier has shown 97\% accuracy in the classification of galaxies, 97\% for stars, and 95\% for AGNs in the 5-dimensional colour space. 
In the test of VIPERS sources with 99\% redshift confidence, the classifier gives an accuracy equal to 94\% for galaxies, 93\% for stars, and 82\% for AGNs. 
The method was applied to sources with low-quality spectra to verify their classification, hence increasing the security of measurements for almost 4~900 objects.
}
{
We conclude that the SVM algorithm trained on a carefully selected sample of galaxies, AGNs, and stars outperforms simple colour-colour selection methods, and can be regarded as a very efficient classification method particularly suitable for modern large surveys.
}

\keywords{Cosmology: observations -- Methods: statistical -- Galaxies: fundamental parameters -- Galaxies: active -- Galaxies: nuclei -- Stars: fundamental parameters}
\maketitle

\section{Introduction}

Over the years, the amount of astronomical data collected by satellites and ground-based surveys is steadily increasing. 
The zoo of collected data, such as photometry, redshifts, spectral lines, and morphology, is constantly expanding, and increasingly researchers are turning to automated algorithms to explore the high-dimensional parameter space.  
Although computationally challenging, the goal is to make use of every available feature to recognise and extract the most discriminating patterns and allow  full systematisation of the data.

Furthermore, the study of the dependence of galaxy properties on physical parameters such as galaxy mass or environment can greatly benefit from the efficient classification of sources.
The classification of different types of sources is one of the basic and, at the same time, crucial tasks to perform before moving on to any scientific analysis. 

The first physical classification of sources in a photometric sky survey is between foreground stars within the Galaxy and extragalactic sources. 
Generally, the distinction between stars and galaxies can be made based upon morphological measurements; point sources are classified as stars, while extended sources are classified as galaxies \citep[e.g. ][]{vasconcellos11,henrion11}. 
For bright apparent magnitudes, the morphology appears to be a reliable criterion for classifying of stars and galaxies, but at fainter magnitudes it becomes difficult to detect low-brightness objects like ultra--compact dwarf (UCD) galaxies, which are often misclassified as foreground stars \citep{drinkwater03}. 
Resolved stellar selection in the current and next generation of wide-field surveys, such as Euclid \citep{laureijis12}, BigBOSS \citep{sholl12}, DES \citep{mohr12}, LSST \citep{ivezic09}, LAMOST \citep{bland12}, and Pan-STARRS \citep{kaiser10}, and/or deep surveys, such as VUDS \citep{lefevre13}, 
HUDF \citep{beckwith06}, DLS \citep{wittman02}, and VISTA \citep{emerson10}, is being challenged by the vast number of unresolved galaxies at faint apparent magnitudes \citep{fadely12}. 
Including of near-infrared photometric bands for many new surveys should improve the classification and separation of faint sources and stars, thereby providing an alternative method of spectroscopy.
 
In the case of fainter sources, colour-colour diagrams are the most widely used tools to separate different classes of celestial sources from one another, since different types of objects will appear in different colour regions in such diagrams due to the shape of the spectral energy distribution (SED).  
For example, galaxies possess  much redder colours than do stars owing to the higher flux at longer wavelengths \citep[e.g.,][]{walker89}. 
Classification methods based on colour-colour selection were employed for star-galaxy separation (e.g.~infrared colour diagram used by \citealp{pollo10}) or for finding special classes of sources, such as  high/low-redshift quasars, active galactic nuclei, starburst galaxies, or variable stars \citep{richards02, stern05, stern12, chiu05, brightman12, wozniak04}.  

Support vector machines (SVMs) are a class of supervised learning algorithms that were created as an extension to nonlinear models of the generalised portrait algorithm developed by Vladimir Vapnik \citep{vapnik95}, for classification in a multidimensional parameter space.   
These algorithms are  based on the concept of decision planes to classify objects using their relative positions in the \textit{n}-dimensional parameter space. 
A large number of observed properties may be analysed simultaneously by the classifier making full use of the data.   
Within the full parameter space, it is possible to build a more reliable classifier than is possible by only using a subset of the data (for example, by analysing only two photometric colours, instead of the complete set). 
On the other hand, the method requires a training sample, that is, a set of data that have known classifications.  
Generally, SVM algorithms are sensitive to the measurement errors and are of limited use for extracting information from noisy data sets \citep{fadely12}. 
The classification of observed sources in astronomy is a fundamental problem, and there is still no approach completely free of drawbacks; however, SVM algorithms are a novel and very promising classification strategy. 

In this paper we apply the SVM algorithm to photometric data. 
Previous works \citep[e.g.,][]{fadely12,solarz12,vasconcellos11,ball06} show high efficiency in that approach for two classes of objects (galaxies and stars). 
Recently, the Photometric Classification Server (PCS) for the prototype of the Panoramic Survey Telescope and Rapid Response System (Pan-STARRS1) based on support vector machines was developed (Saglia et al., 2012). 
The PCS system is using five photometric bands ($\rm{g_{P1}}$, $\rm{r_{P1}}$, $\rm{gi_{P1}}$, $\rm{z_{P1}}$, and $\rm{y_{P1}}$) and is able to separate three groups of sources (stars, galaxies, QSOs) without any preselection based on colour or redshift range and  with high accuracy of galaxy classification ($\sim$97\%). 
The purities of stellar and QSO samples' classifications are worse, at the levels of 85\% and 83\%, respectively. 

We decided to develop a three-class recognition algorithm, which will be able to classify galaxies/AGNs/stars based on the photometric data in The Canada France Hawaii Telescope Legacy Survey (CFHTLS).  
We used, as a training set in colour space objects with the best-quality spectra from the VIMOS Public Extragalactic Redshift Survey (VIPERS) and VIMOS VLT Deep Survey (VVDS) Deep (F02 field) and  Wide (F22 field) data. 
After carefully selecting of objects from VIPERS by SVM and defining characteristic patterns for different types of sources, it will be possible to enlarge the sample of galaxies to be used for more detailed studies. 
We plan to use this trained classifier on a large number of sources possessing low-quality spectra within VIPERS to recover sources that cannot be classified based upon the spectrum alone. 
A majority of objects with lower quality spectral information are absorption line systems with low signal-to-noise ratio. 
Faint red stars and faint passive galaxies are often difficult to distinguish by their spectral features, if the quality of a spectrum is low.
Reconfirmation of a class of such an object by the SVM classifier (galaxy, AGN, or star) based upon the photometric measurements also increases the probability that their spectroscopically measured redshift is correct. 

The paper is organised as follows. 
In Section~\ref{data}, we describe the data used in our analysis, both spectroscopic and photometric. 
Section~\ref{method} describes the principles of the SVM learning algorithm.
In Section~\ref{TS} we introduce the training sample used in our work. 
In Section~\ref{sec_results}, we compare the efficiency of the classifier with and without near infrared data.
Additionally, we present the results of the analysis of the basic tests for the classifiers - self-check and test of the classifier on the VIPERS galaxies with redshift measurements confirmation level equal to 95\%.
The section closes with the selection of the optimal classifier used for our subsequent analysis.  
Section~\ref{test_VIPERS_data} describes the results of our classification of optical near-infrared SVM classifier objects from  the VIPERS samples.  
Finally, in Section~\ref{conclusions} we discuss the advantages and limitations of our current SVM classifier, and we outline our improvements for the presented classifier.

\section{Data}
\label{data}
\subsection{Photometric data}

In this section we present the photometric data used in our work. 
All quoted magnitudes used to develop SVM classifiers are in the AB photometry system and were corrected for foreground Galactic extinction according to the E(B-V) factor derived from Schlegel maps \citep{schlegel98}.
The correction for Galactic extinction was performed for each source separately (see \citealp{fritz13}).
The mean value of E(B-V) factor for the CFHTLS W1 field is equal to 0.02 mag, and for the CFHTL W4 field it is equal to 0.05 mag.

\subsubsection*{CFHTLS photometry}
The CFHTLS, a joint Canadian-French programme, has three distinct survey components: 
(1) the SuperNovae Legacy Survey the ``Deep" survey, (2) the ``Wide" - wide synoptic survey (on which VIPERS survey was based), and  (3) a very wide shallow survey, the ``Very Wide".

The heart of MegaPrime, the wide-field optical imaging facility, is the MegaCam CCD camera \citep{boulade00}.
MegaCam, provides multicolour photometry with  wavelength ($\lambda$) coverage from  3500 to 9400$\rm{\AA}$. 
The main characteristics of the MegaPrime/MegaCam broad band filters are described in Table~\ref{MegaPrime}. 
For a more detailed description we refer the reader to the CFHTLS official web page http://www.cfht.hawaii.edu/Science/CFHTLS/.

The data used in this work are a part of CFHTLS T0005 release \citep{mellier08}, produced at the \texttt{TERAPIX}\footnote{http://terapix.iap.fr/} data centre.
We consider a subsample of CFHTLS T0005 catalogue with spectroscopic redshift measured by VIPERS.

The CFHTLS data are provided in single tiles with effective area of $\sim$ 1 deg square, which partially overlap each other. 
During the preparation of the input data for spectroscopic observations we found the shift in colours between different tiles.
To obtain a homogeneous colour selection of spectroscopic targets, the tile-to-tile correction was performed by using one of the fields overlapping with the VVDS-Deep survey (W1-25) as a representative tile.
The detailed description of the tile-to-tile correction and the explanation of the colour correction method can be found in the survey description paper \citep{guzzo13}.

\begin{table}[ht]
\caption[]{MegaPrime$^*$ and WIRCam$^{**}$ filter characteristics.}
\label{MegaPrime}
\begin{tabular}{p{1.8cm}|l |l |l |l |l|l}
Filter & u$^*$&	 g'&	 r'&	 i'&	 z'  & $\rm{K_{s}}$ \\ \hline \hline
central $\lambda$ (nm)	& 374&	 487&	 628&	 777&	1170 & 2146 \\ \hline
bandwidth (nm)&	 76&	 145&	 122&	 151	& 687 & 325 \\ \hline
max. transmission (\%)&	 77.5&	 93.5&	 96.3&	 98&	 95 & 98 \\ \hline
mag. limit$^{***}$& 25.30 & 25.50 & 24.80 & 24.48 & 23.60 & 22.00 \\ \hline
\multicolumn{7}{l}{\scriptsize{$^*$ http://www.cfht.hawaii.edu/Instruments/Filters/megaprime.html }}\\
\multicolumn{7}{l}{\scriptsize{$^{**}$ http://www.cfht.hawaii.edu/Instruments/Filters/wircam.html/ }}\\
\multicolumn{7}{l}{\scriptsize{$^{***}$ measured as the 50\% of completeness (MegaPrime) and 5$\sigma$ (WIRCam)}}\\
\multicolumn{7}{l}{\scriptsize{for point sources.}}
\end{tabular} 
\end{table}

\subsubsection*{WIRCam data}
In our work, we also used near-infrared $\rm{K_s}$ measurements in the AB magnitude system, which were corrected for galaxy extinction and
taken from Wide-field InfraRed Camera (WIRCam; \citealp{thibault03,puget04}), coming from the dedicated follow-up observations for the VIPERS project (Arnouts et al. 2013, in preparation). 
The $\rm{K_s}$ filter has a central wavelength of 2146nm, and maximum transmission on the level of 98\%. 
One may find the detailed description of WIRCam detector on the WIRCam CFHT web page\footnote{http://www.cfht.hawaii.edu/Instruments/Imaging/WIRCam/}.

\subsection{Spectroscopic data}
\subsubsection*{VIPERS survey}
The VIMOS Public Extragalactic Redshift Survey (see http://vipers.inaf.it) is an ongoing large programme aimed at measuring redshifts for $\sim 10^5$ galaxies at redshift $0.5 < z \lesssim 1.2$, to accurately and robustly measure clustering, the growth of structure (through redshift-space distortions), and galaxy properties at an epoch when the Universe was about half its current age.
The galaxy target sample is selected from optical photometric catalogues  of the Canada-France-Hawaii Telescope Legacy Survey Wide (CFHTLS-Wide, \citealp{goranova09, mellier08}).  
VIPERS covers $\sim24$ deg$^2$ on the sky and is divided into two areas within the W1 and W4 CFHTLS fields.  
Galaxies are selected to a limit of $i_{AB}<22.5$ measured using \texttt{Sextractor}'s \texttt{mag\_auto} \citep{kron80}-like magnitude. 
In addition, a simple and robust colour preselection in $(g-r)$ vs $(r-i)$ is applied to efficiently remove galaxies at $z<0.5$. 
In combination with an efficient observing strategy \citep{scodeggio09}, this allows us to double the galaxy sampling rate in the redshift range of interest with respect to a purely magnitude-limited sample, reaching  an average target sampling rate of $>$ 40\%.    
At the same time, the area and depth of the survey results in a fairly large volume, $5\times10^{7}$~h$^{-3}$~Mpc$^{3}$, analogous to that of the 2dFGRS at $z\sim0.1$ \citep{colless01, colless03}.  
This combination of sampling and depth is quite unique over current redshift surveys at $z>0.5$.

VIPERS spectra are collected with the VIsible imaging Multi-Object Spectrograph (VIMOS, \citealp{lefevre00}) at moderate resolution ($R=210$), using the LR red grism, providing a wavelength coverage of 5500-9500$\rm{\AA}$, for a typical redshift rms error of $\sigma_z$=0.00047 (1+z). 
The full VIPERS area of $\sim 24$~deg$^2$ is covered through a mosaic of 288 VIMOS pointings (192~in the W1 area, and 96 in the W4 area).  
Of the VIPERS spectroscopic targets, more than 51~000~$\rm{K_s}$ counterparts were found: 96\% (80\%) of our spectra for W1 (W4) field have $\rm{K_s}$ measurements. 
More detailed description of WIRCam follow-up survey for VIPERS project can be found in \citet{fritz13} and \cite{davidzon13}.

The redshift quality is quantified at the time of validation by attributing grading flags ($\rm{VIPERS_{Zflag}}$) that are obtained from repeated measurements of redshift for the same sources. 
The $\rm{VIPERS_{Zflag}}$ for galaxies and stars range from a value of 4, indicating $>$99\% of confidence that the measurement is secure, to 0, representing a lack of a reliable estimate of redshift. 
$\rm{VIPERS_{Zflag}}$ equal to nine corresponding to galaxies with only one single clear spectral emission feature. 
Objects classified as AGNs follow the same scheme but their flags are increased by ten. 
A similar system was used and tested for example for VVDS survey \citep{lefevre05}. 
A discussion of the survey data reduction and management infrastructure is presented in \citet{garilli12}. 
An early subset of the spectra used here has been analysed and classified through a principal component analysis (PCA) in \citet{marchetti12}.  
A more complete description of the survey construction, from the definition of the target sample to the actual spectra and redshift measurements, is given in the parallel survey description paper, \citet{guzzo13}.

The data set used in this paper are those of the early science data release of VIPERS data as described in \citet{guzzo13}; see also \citet{delatorre13}, \citet{fritz13}, \citet{marulli13}, \citet{bel13}, and \citet{davidzon13}.
This data will be publicly available in fall 2013 as the VIPERS Public Data Release 1 (PDR-1) catalogue.  
This catalogue includes $55,358$ redshifts and corresponds to the reduced data as it was in the VIPERS database at the end of the 2011/2012 observing campaign.

Using the automatic source classifier for VIPERS data is a  natural step to handle this unique data volume.
Automated and efficient source classifiers based on photometric observations, can provide class labels for catalogues and  be used to recover objects for study according to various criteria.
Moreover, a multilevel SVM classifier, trained to search for specific types of sources such as active galactic nuclei (AGNs) or galaxies, with an additional redshift measurement  as a feature in the parameter space, can be used to boost confidence in the reliability of redshift estimates for sources with poor spectroscopic data. 
We are planning to develop a more sophisticated and detailed classifier in the near future, enlarging the parameter space by  adding measurements of spectral lines and galaxy morphological parameters, thus enabling a finer classification of our sources (e.g. distinguish among different galaxy types).  

In this work, we used VIPERS data both to construct a training sample and to select samples on which to apply the classifier to separate three different classes of objects (galaxy/AGN/star). 

\subsubsection*{VIMOS-VLT Deep Survey (VVDS)}
VIPERS was designed as an extragalactic survey that aims to efficiently measure of redshifts for a large sample of galaxies. 
To increase the efficiency, stars were carefully removed from the target candidates (which was particularly important for the W4 VIPERS field owing to its low galactic latitude).
To this aim, both morphological and spectral energy distribution fitting techniques were used \citep[see ][]{guzzo13, coupon09}. 
However, it was also important to re-introduce AGNs, which were identified among the stellar objects by their photometric properties \citep[a more detailed description of AGN selection can be found in the survey description paper, ][]{guzzo13}.  
Consequently, the number of observed stars and AGNs in VIPERS is quite small. 

To construct a reliable training sample (see Sect.~\ref{method}), we included data from another, similar, but more complete survey, VVDS. 
The VVDS fields, like VIPERS, are covered by CFHTLS (and partially by WIRCam observations) and thus the photometric information is homogeneous.  
Additionally, both surveys utilise the VIMOS spectrograph in similar configurations. 
The VVDS spectroscopic sample is based upon a purely magnitude-limited selection such that the survey contains a much wider variety of sources than VIPERS. 
We used VVDS-Deep (F02 field)  and VVDS-Wide (F22 field)  surveys to construct a training sample  of AGNs (objects classified as AGNs by \citealp{gavignaud07}).  
The stellar sample was chosen from a part of VVDS Wide F22 that overlaps the VIPERS W4 field.

The Deep F02 survey, covering 0.49 square degrees, is a purely magnitude limited sample to $\rm{i_{AB}}\mbox{ }\leq$ 24. 
The detailed description of the VVDS Deep survey may be found in  \citet{lefevre05}. 
The VVDS Wide F22 survey \citep{garilli08}, covering an effective area three square degrees, is also a magnitude limited survey with limitation to $\rm{i_{AB}}$=22.5. 

\section{Method - support vector machines}
\label{method}
The main purpose of the support vector machine (SVM) is to calculate decision planes between a set of objects having different class memberships.
A so-called training sample, a training set of objects, is used to provide the SVM with examples of the different classes of sources.  
The SVM searches for the optimal separating hyperplane between the \textit{n} different classes of objects by maximising the margin between the classes closest points (the so-called support vectors).
Instead of using the probability function as in Bayesian statistics or template-fitting methods, the objects are classified based on their relative position in the \textit{n}-dimensional parameter space with respect to the separation boundary. 
A well chosen training sample is at the heart of the method, because, based on the properties of the training sample, the classifier is tuned, and the hyperspace between classes is determined. 
\begin{figure}[t]
 \begin{center}
	\resizebox{0.9\hsize}{!}{\includegraphics{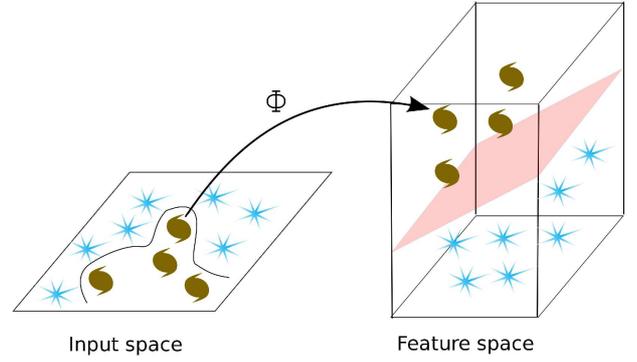}}
	\caption[]{An illustration of the operation of the SVM algorithm. 
	The input data (on the left side) are transformed by a kernel into the higher dimensional feature space (right side) 
	where, instead of having a complex boundary separating different classes of objects,
	we can find an optimal separating hyperplane.	
	}
	\label{psi}
 \end{center}
\end{figure}

The SVM algorithm represents a major development in machine-learning techniques.  
It can be applied to classification or regression problems and is nowadays  constantly growing in popularity, to deal with astronomical  data for distinguishing different classes of sources based on a multidimensional space of parameters taken from observations. 
Recently, \citet{wozniak04} has used SVMs efficiently to analyse variable sources in a five-dimensional space constructed from the period, amplitude, and three colours. 
\citet{huertas08} quantified the morphologies of near-infrared galaxies based on 12-dimensional space, including five morphological parameters and other characteristics of galaxies, such as luminosity and redshift.  
\citet{solarz12} created a star-galaxy separation algorithm based on mid and near-infrared colours, and 
\citet{saglia12} separated three different classes of sources (galaxies, QSOs, and stars) from the PAN-STARRS1 survey, based on five photometric bands.
Last year brought a significant number of astronomical papers that implement supervised machine-learning algorithms to handle various tasks, not only to classify sources but also to predict characteristic features of specific objects. 
For example,  \cite{peng12} used SVM to select AGN candidates and to estimate redshift, \cite{hassan13} - to search specific AGN subclass: BL Lacertae and flat-spectrum radio quasars based on the  \textit{Second Fermi LAT Catalogue}). 
Clearly SVMs present an innovative method with great potential to be widely used in many different branches of astronomy, a potential we are just beginning to tap into.

We used the SVM algorithm to build a non-linear classifier for photometric data to select three different classes of objects: galaxies, AGNs, and stars. 
The first step in our classification task involves selecting a secure training sample of galaxies, AGNs, and stars, taking advantage of the redshift information provided by VIPERS and VVDS and using their attributes - i.e. their observed photometric fluxes - to train the SVM.  

The algorithm, aided by a non-linear kernel function, searches for a hyperplane that will maximise the distance from the boundary to the closest points belonging to the separate classes of objects \citep{crisrianini00, shawe04}.
The kernel is a symmetric function $\Phi$ that maps $\rm{k : X \times X \rightarrow F}$, so that for all  $\rm{x_i}$ and $\rm{x_j}$,  $\rm{k(x_i,x_j) = <\Phi(x_i),\Phi(x_j)>}$ from the input space X to the feature space F \citep{vanschoenwinkel05}, see Fig.~\ref{psi}. 
For our analysis we chose a Gaussian radial basis kernel (RBK) function, defined as  
\begin{equation}
\label{Gaussian}
 \rm{k(x_i,x_j)=\exp{(-\gamma||x_i-x_j||^2})},
\end{equation}
where $\rm{||x_i-x_j||}$ is the Euclidean distance between $\rm{x_i}$, and $\rm{x_j}$. 
The effect of the kernel function  is a non-linear representation of each parameter from the input to the feature space. 
The RBK kernel is one of the most popular SVM kernel functions, used to make the non-linear feature map. 
We decided to use it because of its effectiveness and  simplicity in adjusting the free parameters.

\begin{figure}[t]
 \begin{center}
	\resizebox{0.82\hsize}{!}{\includegraphics{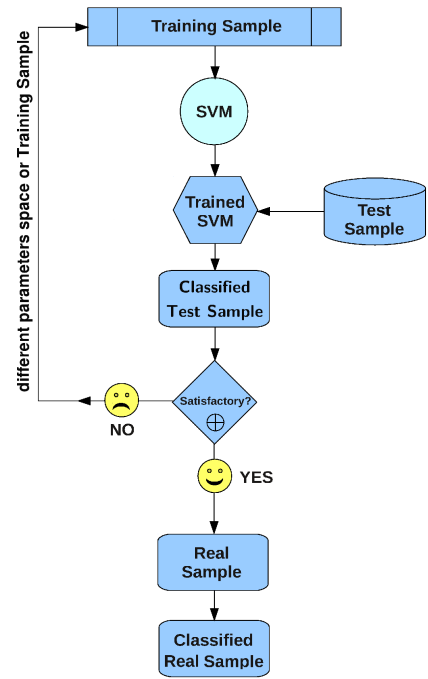}}
	\caption[]{A schematic representation of the SVM algorithm classification process.
	We take as input the preselected training sample consisting of (in the case of this work) three distinct classes of objects. 
The SVM is taught how to distinguish one class from the others based on the discriminating properties chosen as feature vectors.
Then, the classifier is trained by tuning the free parameters (\textit{C} and $\gamma$).
If the result reaches a high enough accuracy rate (the number of objects from the training sample that are correctly recognised by the classifier) 
without overfitting (the resulting hyperplane does not confine the sources of a specific type too tightly), it will be used to classify the unknown objects (test sample). 
If the accuracy is not satisfactory, a different parameter space (or training sample, if possible) is chosen to tune \textit{C} and $\gamma$. 
After a number of iterations, which allow the classifier to reach high enough efficiency level, a real sample can be classified using the discriminant hyperplanes.
	}
\label{SVMchart}
 \end{center}
\end{figure}

For our tasks,  we used a soft-boundary SVM method called \textit{C}-SVM.
We chose \textit{C}-classification because of its good  performance and only two free parameters: 
\begin{itemize}
 \item \textit{C} - a trade-off parameter that sets the width of the margin separating different classes of objects. 
  A large  \textit{C} value sets a small margin of separation between different classes of objects; however increasing the \textit{C} parameter too much can lead to over-fitting.  
  Reducing \textit{C} will make the hyperplane between different classes of objects smoother, allowing for some misclassifications.
  \item $\gamma>0$ parameter (related to the kernel function) determines the topology of the decision surface.
   A low value of $\gamma$ sets a very rigid, and complicated decision boundary; a value of $\gamma$ that is too high can give a very smooth decision surface causing misclassifications.
\end{itemize}
A schematic representation of the SVM algorithm classification process, beginning with choosing the training sample, tuning C and $\gamma$ parameters, self-checking of the classifier, and finally, classifying the real sample is shown in Fig.~\ref{SVMchart}.

For our analysis we used LIBSVM\footnote{{http://www.csie.ntu.edu.tw/cjlin/$\sim$libsvm/}} \citep{chang11}, an integrated software for support vector classification, which allows for multiclass classification.
We used R\footnote{http://www.r-project.org/}, a free software environment for statistical computing and graphics, with e1071 interface \citep{meyer01}  package installed.

\section{Training sample}
\begin{figure*}[ht]
 \begin{center}
	\resizebox{1\hsize}{!}{\includegraphics{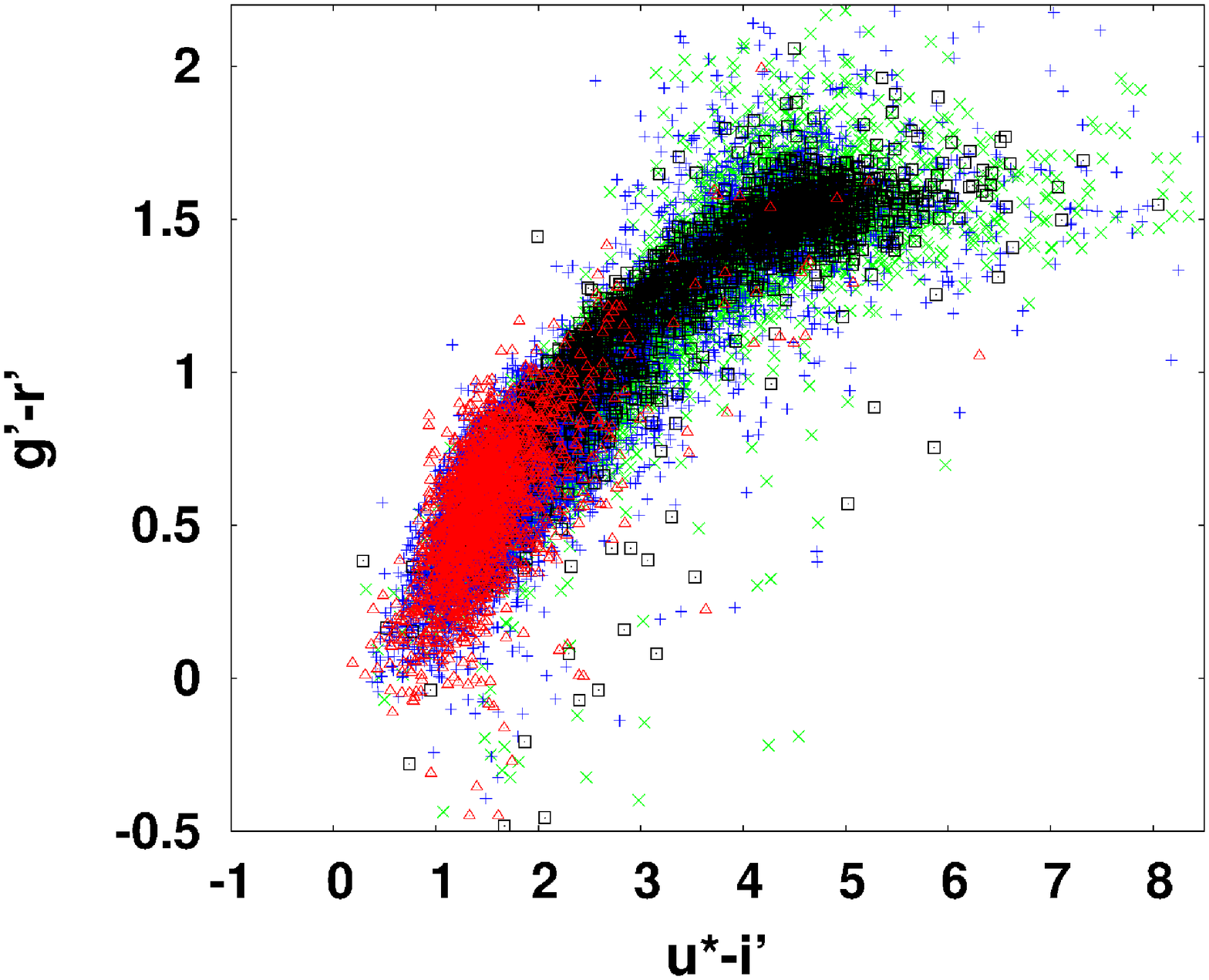}\includegraphics{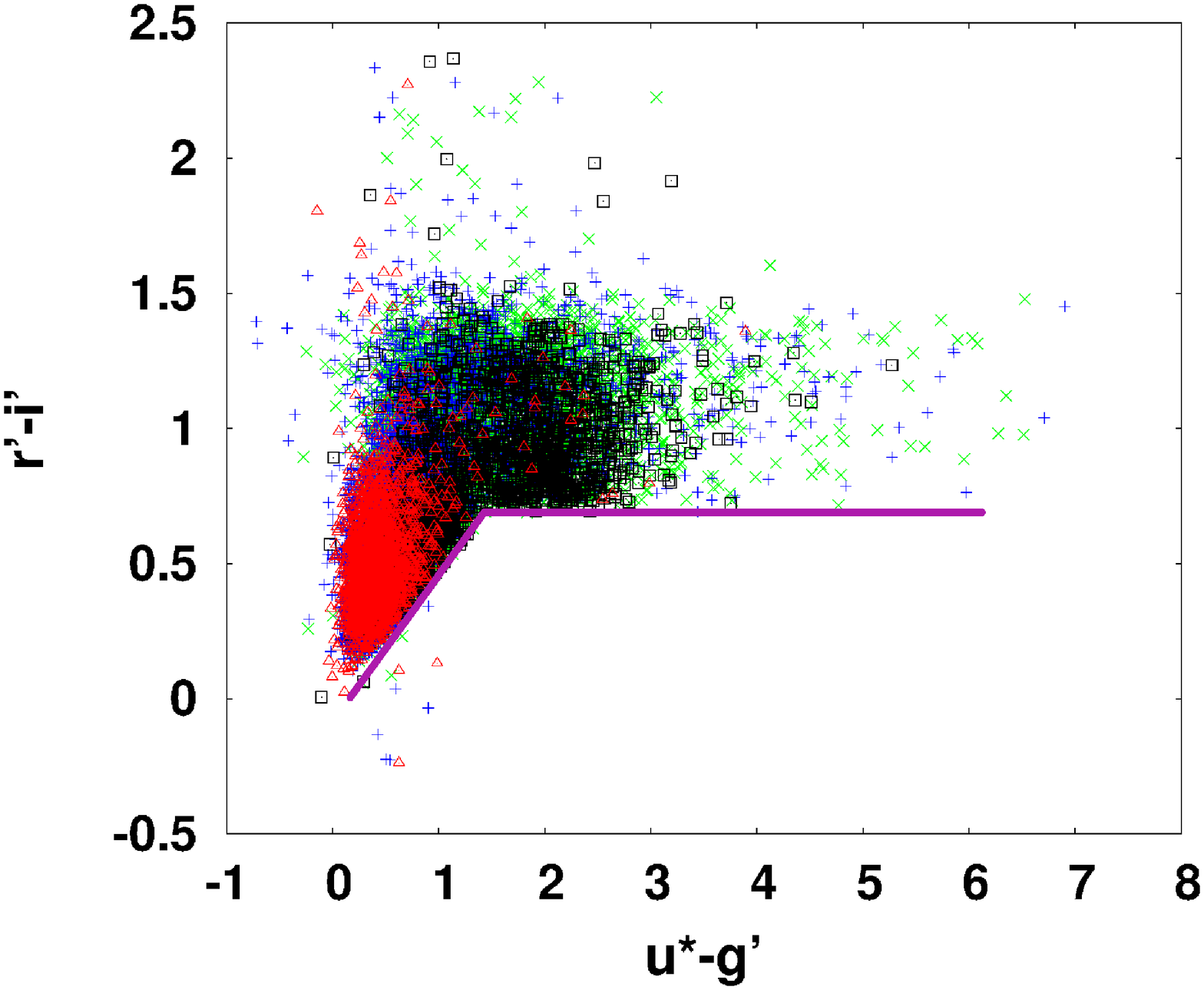}\includegraphics{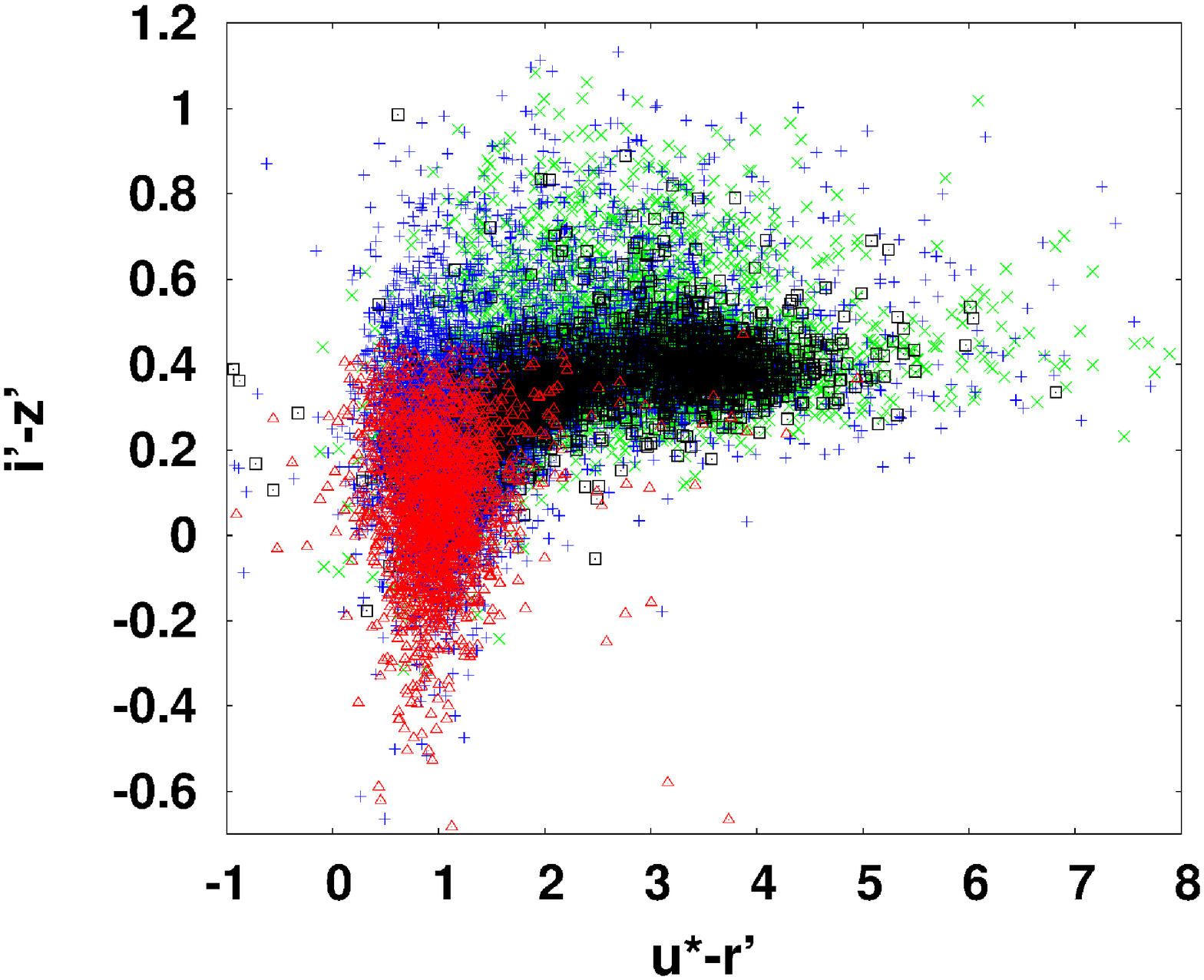}}
	\caption[]{The representative  colour-colour plots for the galaxy training sample.
	Open black squares represent objects  with i'-apparent magnitude between 19 and 20 mag; 
	green \texttt{X}-s - galaxies  with i' magnitude between 20 and 21 mag; 
	objects with i' apparent magnitude between 21$\leqslant$i$<$22, and 22$\leqslant$i$<$22.5 mag are marked as blue \texttt{+}-s and open red triangles, respectively; 
	In the middle panel of colour-colour plots, the boundaries of VIPERS selection are marked as magenta lines.
	}
\label{GALs4}
 \end{center}
\end{figure*}

\begin{figure*}[ht]
 \begin{center}
	\resizebox{1\hsize}{!}{\includegraphics{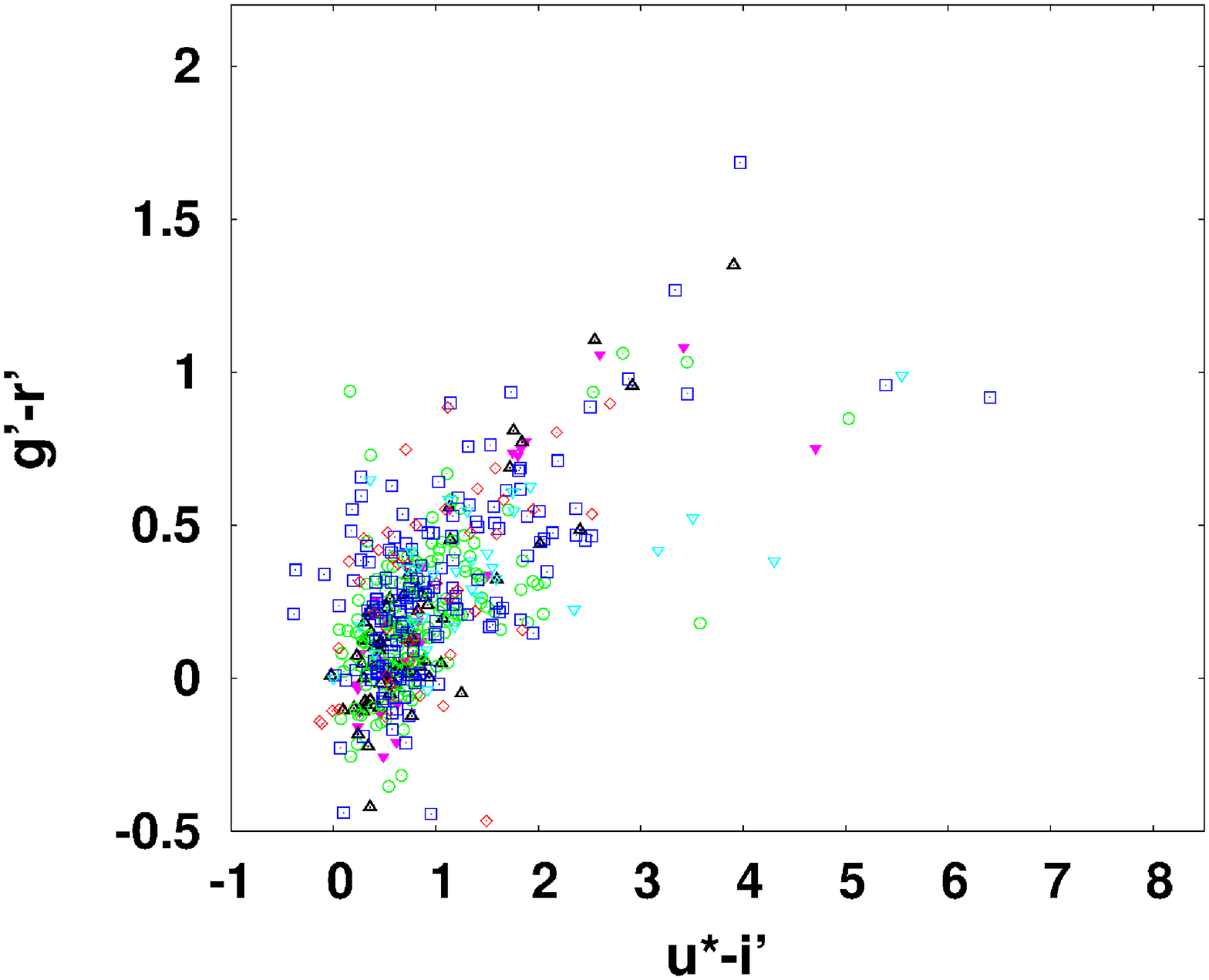}\includegraphics{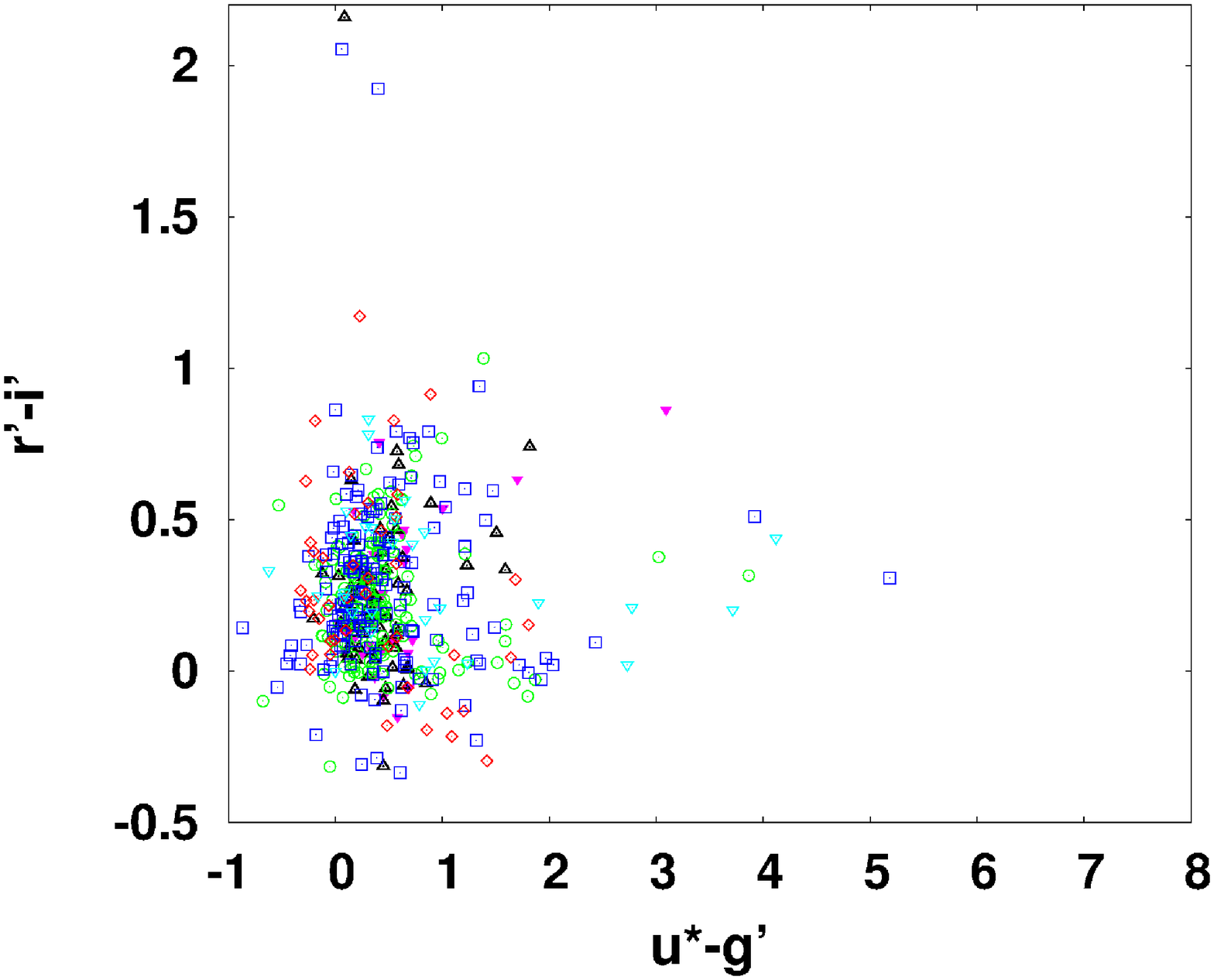}\includegraphics{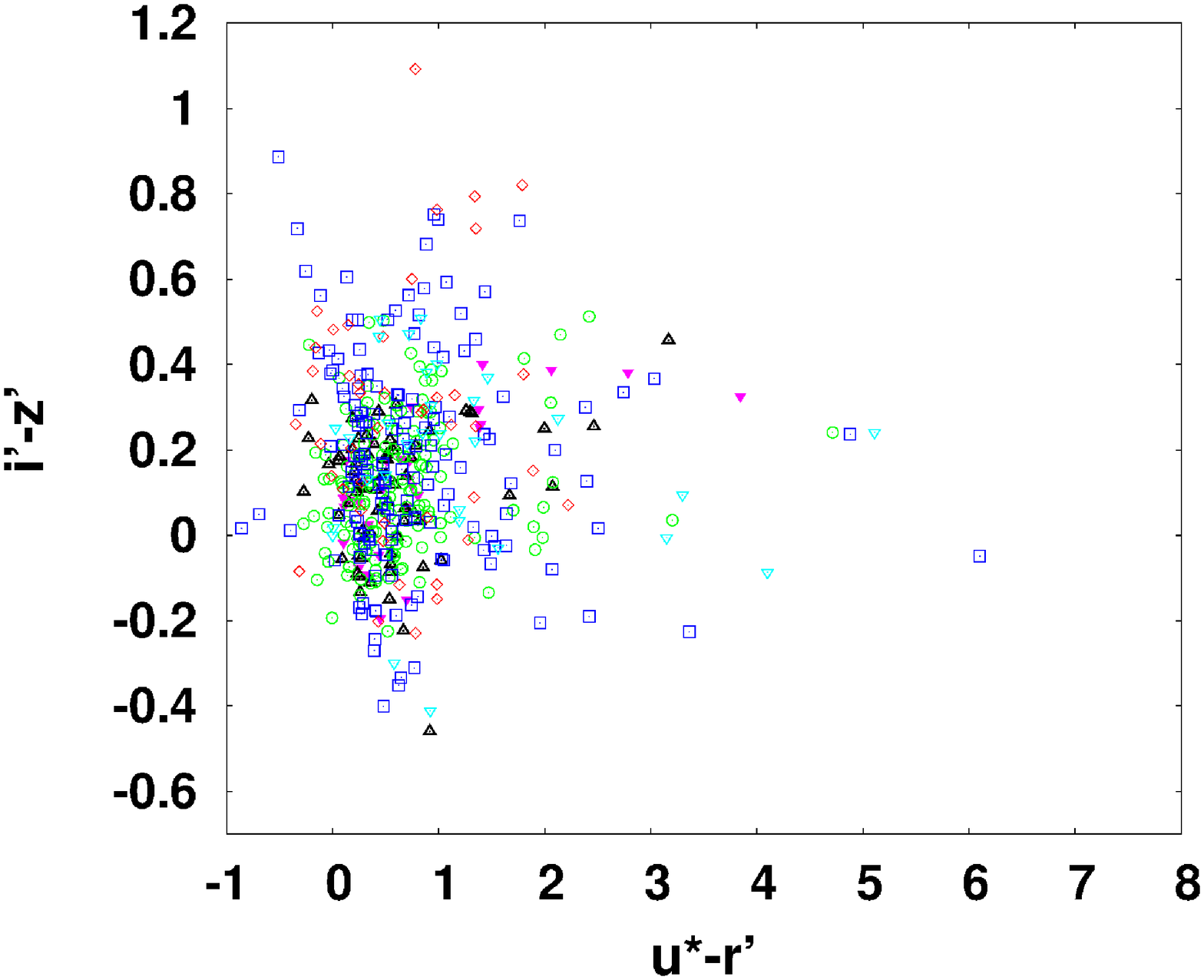}}
	\caption[]{The representative  colour-colour plots for the AGN training sample.
	Full magenta triangles represent objects brighter than 19 mag in the i' band.
	Open black triangles - AGNs with i-apparent magnitude between 19 and 20 mag; 
	open green circles - AGNs  with i' magnitude between 20 and 21 mag;
	objects with i' apparent magnitude between 21$\leqslant$i$<$22, and 22$\leqslant$i$<$22.5 mag are marked as open blue squares and open red diamonds, respectively;
	AGNs with i' apparent magnitude fainter than 22.5 are marked as open rotated cyan triangles.
	}
\label{AGNs}
 \end{center}
\end{figure*}

\begin{figure*}[ht]
 \begin{center}
	\resizebox{1\hsize}{!}{\includegraphics{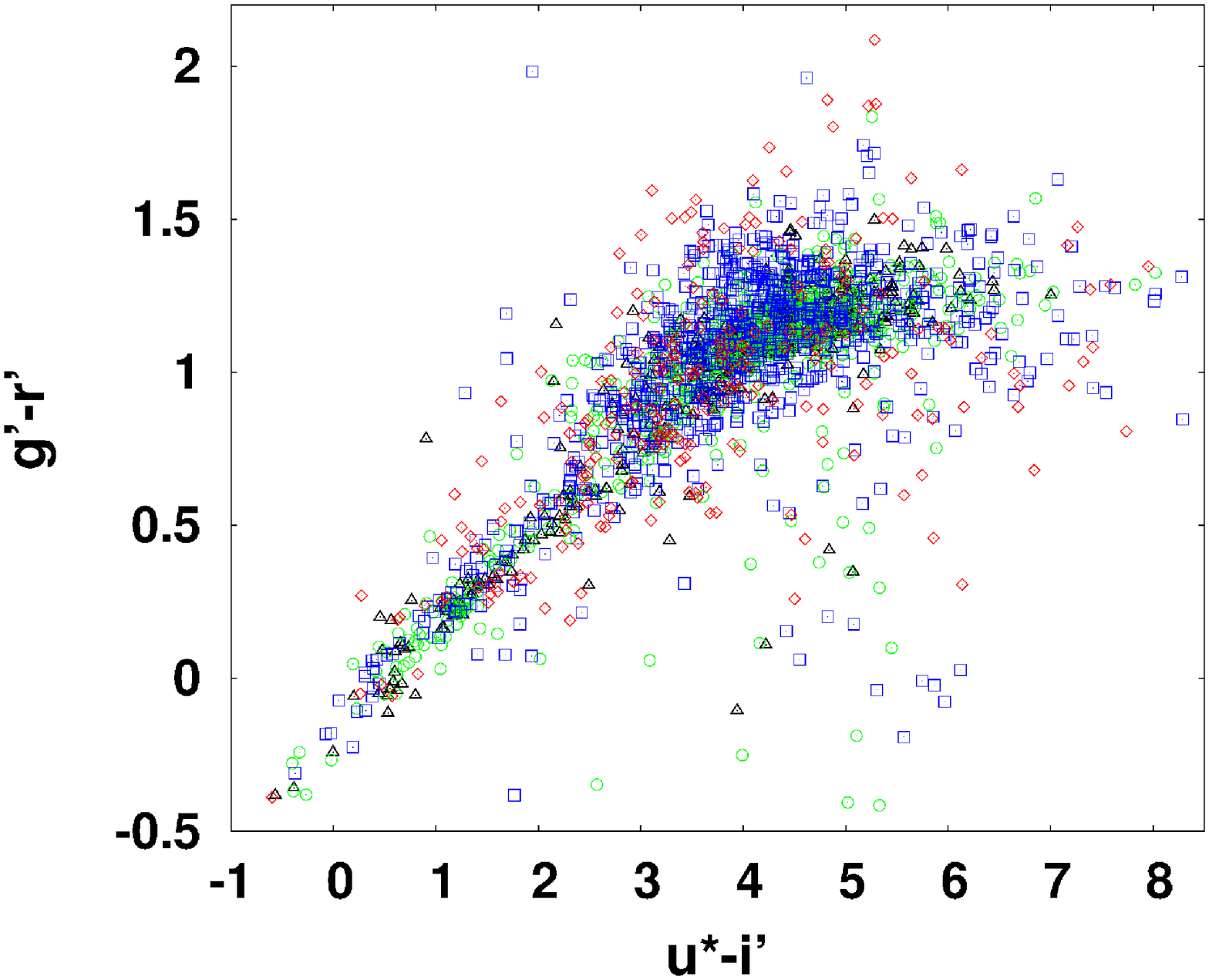}\includegraphics{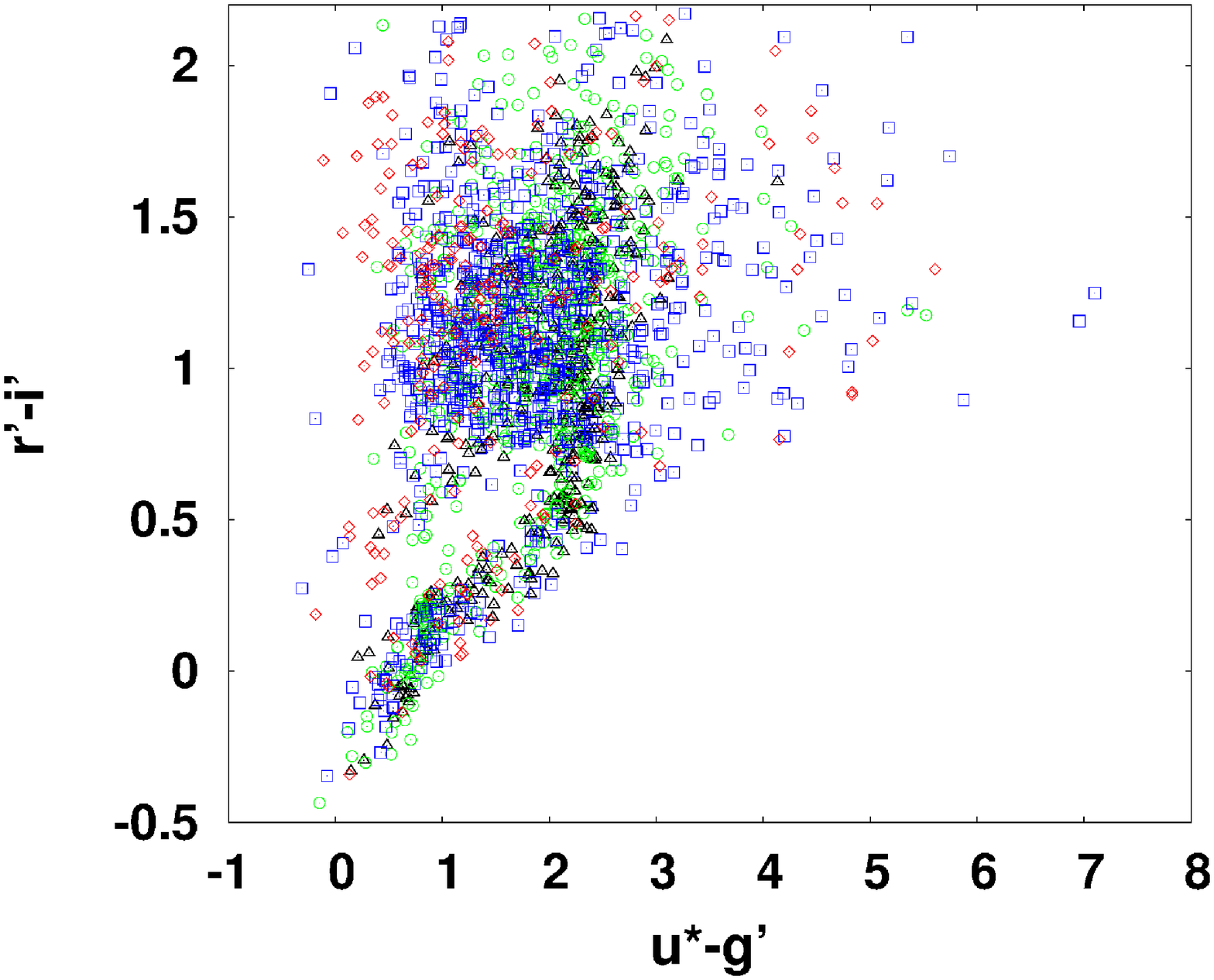}\includegraphics{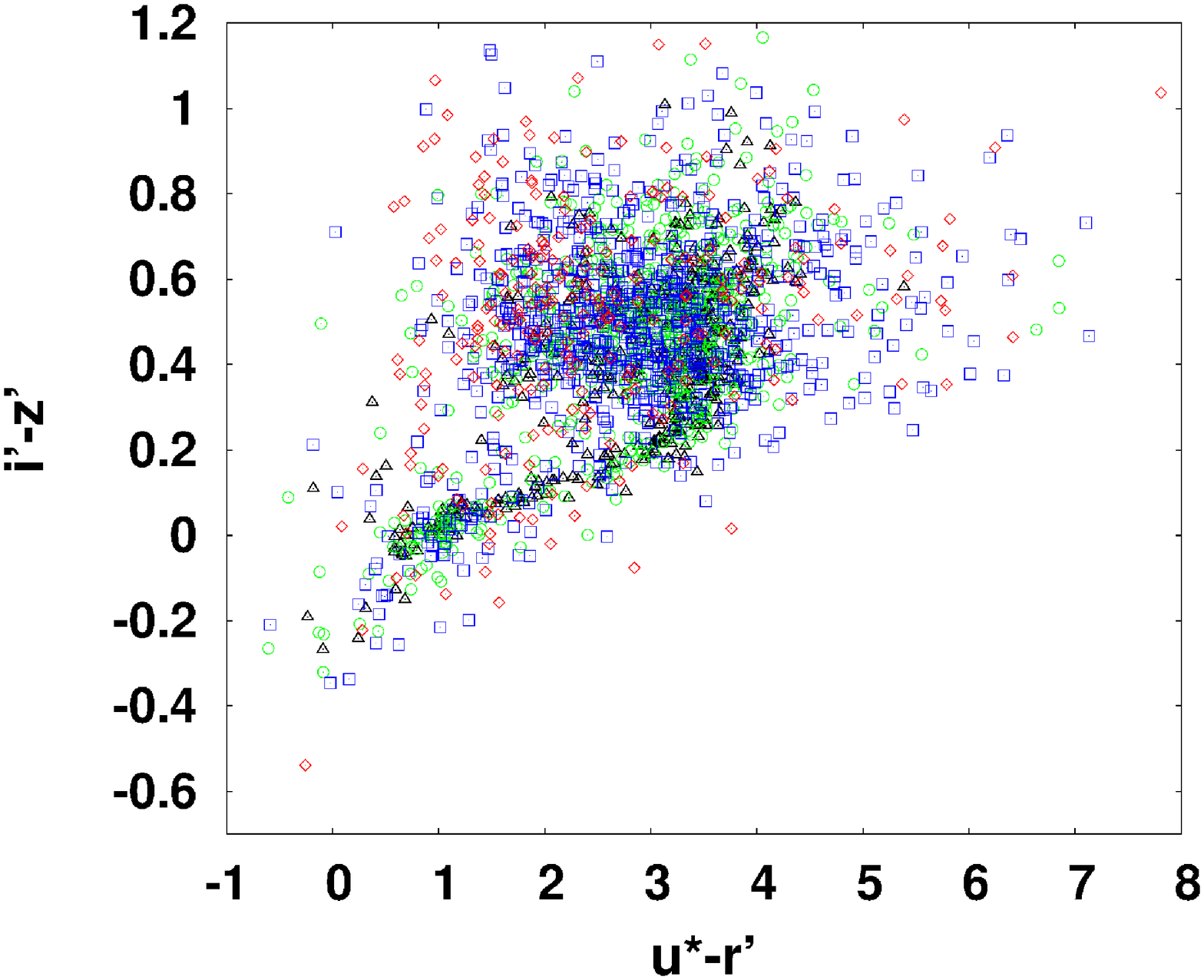}}
	\caption[]{The representative  colour-colour plots for the star training sample.
	Open black triangles - stars with i-apparent magnitude between 19 and 20 mag; 
	open green circles - stars  with i' magnitude between 20 and 21 mag;
	objects with i' apparent magnitude between 21$\leqslant$i$<$22, and 22$\leqslant$i$<$22.5 mag are marked as open blue squares and open red diamonds, respectively.
	}
\label{STARS}
 \end{center}
\end{figure*}

\label{TS}
The successful application of an SVM algorithm requires a carefully selected training sample - a set of objects with confirmed classes which will serve as a template for distinguishing the sources whose class we want to determine.
Since this work is focused on the selection of galaxies, AGNs, and stars we select as a training sample a set of sources whose basic class (galaxy, AGN or star) was established with the highest reliability thanks to their high quality spectra (their redshift being measured with the highest confidence flag within the VIPERS or VVDS surveys). 
For these sources the accurate photometric information provided by the CFHTLS wide-survey and the WIRCam follow-up observations of the VIPERS/VVDS fields, provided the colour information needed to create the discriminant vectors for training our SVM algorithm. 
We produced a model (the optimised $C$ and $\gamma$ parameters based on the training data), which predicts the target values of the test data given only the test data attributes \citep{hsu10}.

\subsection{Galaxies}
As a galaxy training sample we used the sources with the best redshift measurements in both the W1 and W4 VIPERS fields ($\rm{VIPERS_{Zflag}}$~=~4, corresponding to the highest confidence level of redshift measurements and thus of spectroscopic classification as a galaxy). 
It is useful to remember that VIPERS is preselected not only in magnitude (i'$<$22.5) but also in colours: (r'-i')$>$0.5*(u$^*$-g) or (r'-i')$>$0.7. 
We have divided the galaxy training set into i'-based apparent magnitude-binned samples and trained the classifier on each subset.
As a galaxy training sample we used 16~271 galaxies: 1~884, 5~483, 6~778, and 3~226 for 19$\leqslant$i'$<$20, 20$\leqslant$i'$<$21, 21$\leqslant$i'$<$22, and 22$\leqslant$i'$<$22.5 apparent magnitude-bins, respectively.
Based on our initial tests, we decided to divide our galaxy sample into the magnitude bins to separate more efficiently  different groups of galaxies seen in  different i' apparent magnitude ranges  to improve their classification. 
Figure~\ref{GALs4} shows that galaxies in different magnitude bins occupy different areas of the colour-colour plots, partly because of different redshift range and different morphology. 

\subsection{AGNs}
\label{AGNsTS}
Given the small number of AGNs detected in the VIPERS fields with the $\rm{VIPERS_{Zflag}}$ = 14, we increased the AGN sample by using all AGNs which had at least 99\% confidence level of spectroscopic 
classification ($\rm{VIPERS_{Zflag}}$ 13 and 14, in total 398 objects). 
AGN spectra are quite easy to recognise, so a lower flag on the quality of the measured redshift  does not infringe on the reliability  of the classification as an AGN.  
There are two ways that an AGN can be observed in VIPERS: 
\begin{itemize}
\item It is star-like and meets the AGN candidate selection.
This includes samples of X-ray selected AGNs from the XMM-LSS survey, overlapping the VIPERS W1 field \citep{pierre04}, and  AGNs selected by colour-colour criteria from the sample of star-like sources that would otherwise not be targeted.  
\item It meets the galaxy selection criteria - AGNs which met the galaxy criteria during the main VIPERS colour preselection. 
\end{itemize}

We stress that the colour preselection for galaxies and AGNs is slightly different, and AGNs occupy only a part of the full colour-colour galaxy plane. 
The first AGN colour separation criterion $\rm{CC_{1AGN}}$:

\begin{equation}
\label{cc1}
 \rm{(g'-r')<1} \wedge \left\{ 
    \begin{array}{l}
 
    1.\mbox{ }   \rm{(u^*-g)_{corr}<0.6,} \\
    2.\mbox{ }   \begin{array}{l}
		  \rm{0.6 \leqslant (u^*-g)_{corr}<1.2}\mbox{ }\&\\ 
		  \rm{(g'-r')_{corr}>0.5(u^*-g')_{corr}+0.036,}
		  \end{array}\\
    3.\mbox{ }    \begin{array}{l}
		  \rm{0.6 \leqslant (u^*-g)_{corr}<2.6}\mbox{ }\&\\ 
		  \rm{(g'-r')_{corr}<0.5(u^*-g')_{corr}+0.214,}
		  \end{array} \\
    4.\mbox{ }    \rm{(u^*-g')_{corr}>2.6,}
   \end{array}
 \right.
 \end{equation}
where $\rm{(u^*-g)_{corr}}$ and $\rm{(g'-r')_{corr}}$ correspond to tile colour offset.

\begin{figure}
 \begin{center}
	\resizebox{0.9\hsize}{!}{\includegraphics{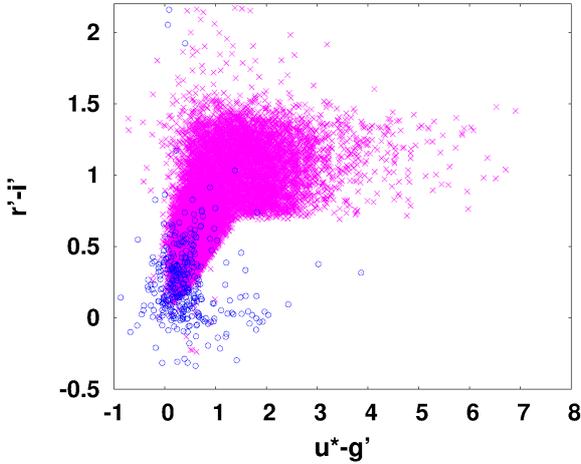}}
	\caption[]{The representative colour-colour plot for VIPERS galaxies with $\rm{VIPERS_{Zflag}}$ = 4 (pink \texttt{x}-s) and AGNs with $\rm{VIPERS_{Zflag}}$ = 3 and 4 (open blue circles).	
	}
\label{GALAGN}
 \end{center}
\end{figure}

The colour-colour selection criterion of AGNs, given in Eq.~\ref{cc1}, was based on the results from the VVDS survey.
After one year of observations it turned out that this selection  criterion introduces a stellar contamination  at the level $\sim$60\%.   
From August 2010, additional criterion $\rm{CC_{2AGN}}$, including  the (g'-i') vs (u$^*$-g') colour-colour plane, was added to eliminate stellar sample from AGNs targets.
The set of colour-colour criteria included to $\rm{CC_{2AGN}}$ is
\begin{equation}
\label{cc2}
  \left\{ 
    \begin{array}{l}
    1.\mbox{ }\rm{(u^*-g')_{corr}<0.6}\mbox{ } \&\mbox{ } \rm{-0.2<(g'-i')<1,} \\
    2.\mbox{ }\begin{array}{l} \rm{0.6\leqslant(u^*-g')_{corr}<1}\mbox{ } \&\\{ }\rm{-0.2<(g'-i')<0.2,}\end{array} \\
    3.\mbox{ }\rm{(u^*-g')_{corr}\geqslant1\mbox{ } \&\mbox{ }(g'-i')<0.6.}
  \end{array}
   \right.
\end{equation}
Therefore, both criteria (Eqs.~\ref{cc1} and ~\ref{cc2}) applied simultaneously defined VIPERS AGN targets.
However, most of the AGNs share the same colour-colour space as galaxies (as can be seen in Fig.~\ref{GALAGN}).
A part of AGNs occupy different colour-colour areas than galaxies and for them, the galaxy/AGN separation is not so difficult. 
For objects classified as AGNs lying in the same colour-colour plane, the galaxy/AGN/star separation is more challenging. 
For this reason we decided to use SVM with \textit{n}-dimensional photometric parameter space to classify sources with similar properties in the typical colour-colour plane. 
That is why it is a challenge to distinguish all three classes of objects using an automatic classifier.

To enlarge the AGN training sample, we also merged the VIPERS sample with objects classified as broad-line AGNs in the VVDS survey.
In our training sample we included AGNs identified by \citet{gavignaud07} - a catalogue of broad emission-line AGNs, from the purely flux-limited spectroscopic sample of the VVDS survey. 
No colour-based preselection has been applied to these AGNs.
For our studies we used 100 AGNs from VVDS Deep F02 \citep{lefevre05} and VVDS Wide F22  \citep{garilli08} fields only. 
We selected these fields since they have the same CFHTLS photometry system as the VIPERS survey. 
We found that AGNs detected in both VIPERS fields do not display any systematic difference in the colour-colour distribution, confirming that our extinction correction works well. 

Cumulatively, our AGN training sample reached 498 objects. 
A part of them, observed by VIPERS, preselected by colour. 
AGNs from VVDS fields have no colour preselection (flux-limited only).
Since we checked on colour-colour plots (see Fig.~\ref{AGNs}), in the different magnitude bins, we do not see a change in population of our AGN sample with apparent luminosity. 
For this reason, unlike the case of the galaxy sample, we decided not to divide the AGN training sample into i'-based apparent magnitude binned samples, but to use it as a whole in each bin to increase the population of the training AGNs.

\subsection{Stars}
VIPERS performed a star/galaxy classification in the CFHTLS wide fields to effectively remove stars from the sample of observed targets. 
This procedure is crucial, since at i'$<$22.5 the fraction of stars can be as high as 50\% (as in the case of W4, \citealp{guzzo13}). 
The basic VIPERS classification procedure was based on the colour-colour preselection with $\rm{(r-i)>0.5*(u-g)}$ or $\rm{(r-i)>0.7}$, but owing to the low galactic latitude of W4 field, VIPERS implemented an additional procedure.
We refer the interested reader to \citet{guzzo13} for a complete description of the adopted strategy, but here it is sufficient to mention that 
for objects brighter than i'=21 an additional preselection based  on the observed angular size of sources was applied, 
while for objects fainter than i'=21 a combined method making use of an angular size and SED fitting by the Le Phare code \citep{arnouts99,ilbert06} has been used.  
These preselection criteria proved to be very effective.
However, the average stellar contamination in the VIPERS database, for both fields, remains on the level of 3.2\% (1.49\% and 4.86\% for the W1 and W4 fields, respectively).
It means that in the VIPERS PDR-1 catalogue, which includes 55~358 objects, only 1~750 objects have been identified as stars. 
In sum, the VIPERS PDR-1 catalogue contains 1~750 (3.20\%) stars classified as galaxies in the beginning, with colours compatible with an object at z$>$0.5. 
This stellar sample can be divided into two main groups: 
\begin{itemize}
 \item stars that were not distinguishable from galaxies based on the VIPERS preselection criteria, and 
 \item stars that were included in the sample as AGN candidates.
\end{itemize}
Then, it should be stressed that the stars observed by VIPERS are  interlopers within the galaxy and AGN samples and are thus  not representative of the stellar class. 
However, our method uses the multidimensional colour space which opens a possibility that in such a space, these sources may occupy a region  separated from galaxies and AGNs. 

\begin{figure*}[ht]
 \begin{center}
	\resizebox{1\hsize}{!}{\includegraphics{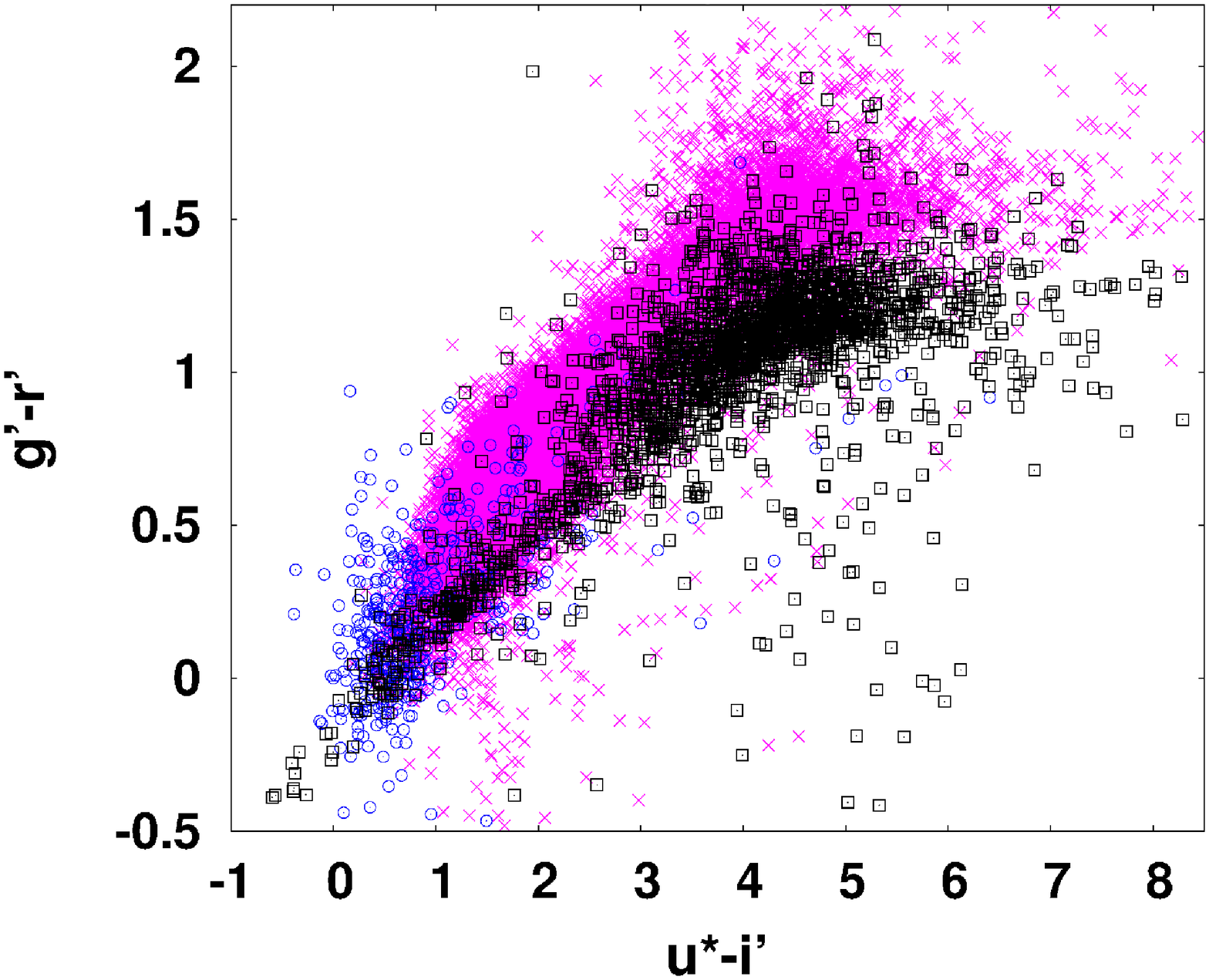}\includegraphics{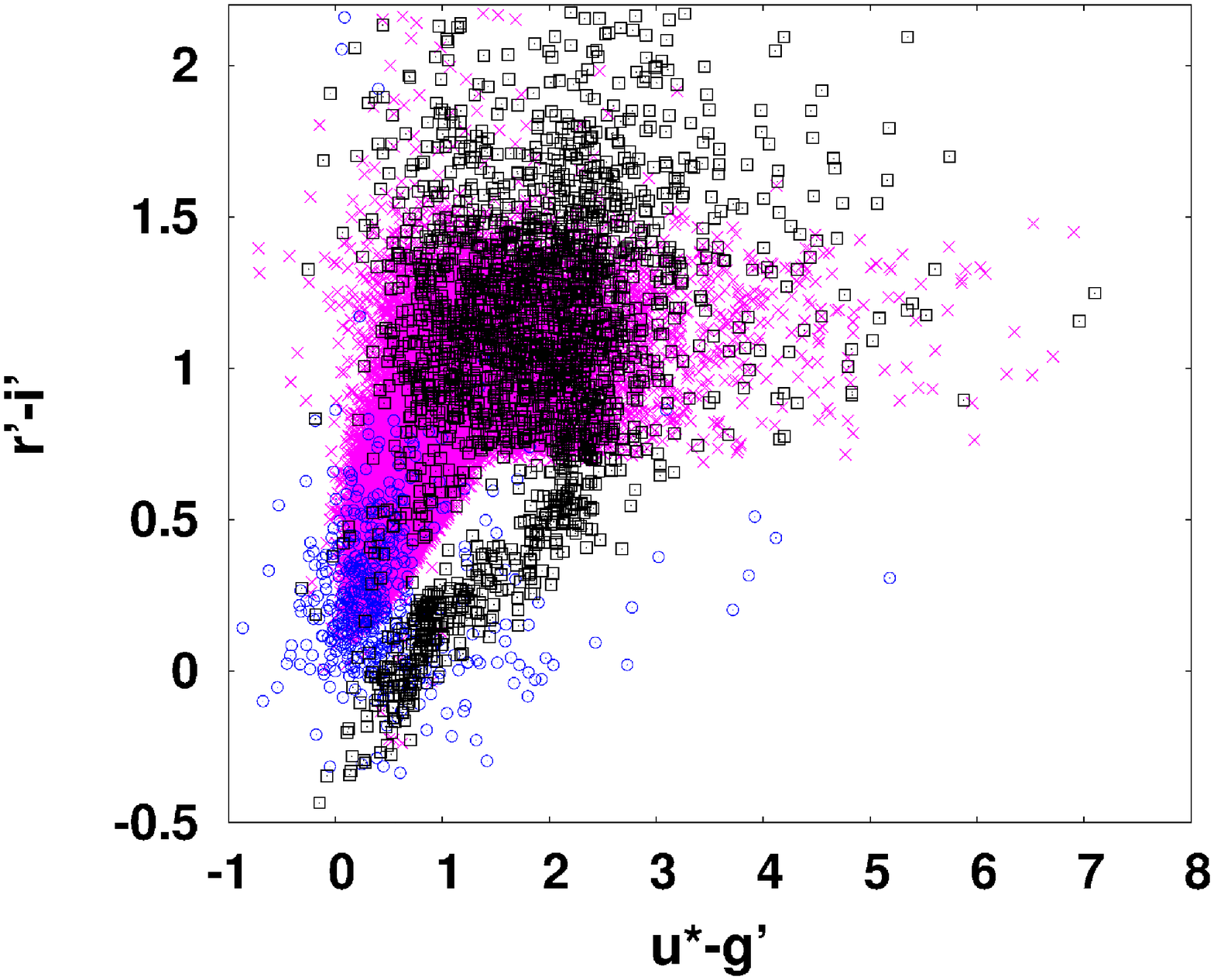}\includegraphics{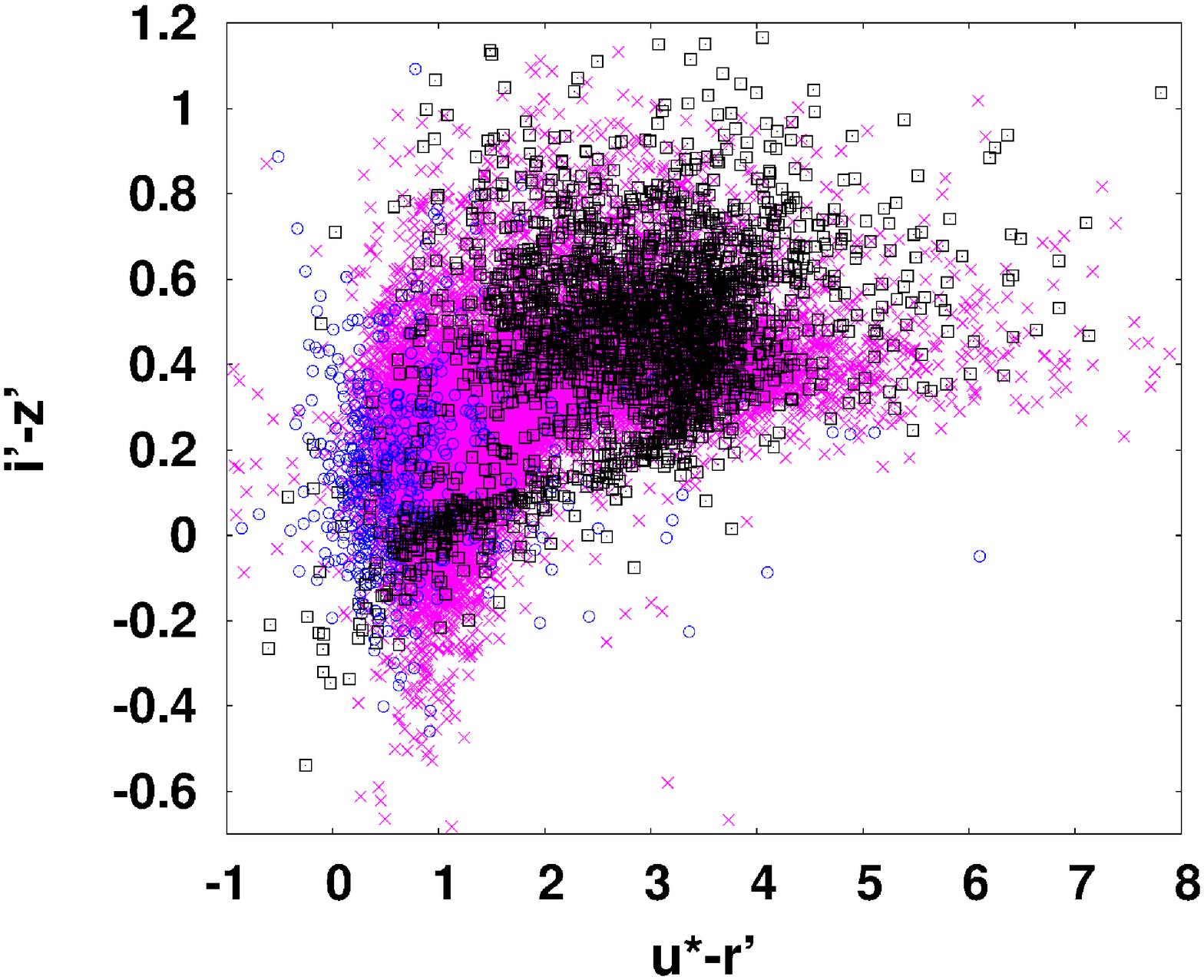}}
	\caption[]{The representative colour-colour plots for all objects used for the training sample.
	Pink \texttt{x}-s represent galaxies. 
	Open blue circles correspond to the AGN sample, and open black squares to the stellar sample.}
\label{ALL}
 \end{center}
\end{figure*}

To build an unbiased star training sample we added spectroscopically classified stars from the VVDS Wide F22 overlap with the VIPERS W4 field.
VVDS Wide F22 observations were carried out on the same magnitude limits sample as VIPERS, but without any photometric preselection.  
The overlap between the VVDS Wide F22 and VIPERS W4 fields contains 920 objects spectroscopically classified as stars by VVDS in the 19$\leqslant$i'$<$22.5 apparent magnitude bin. 
We increased the stellar training sample by using all VIPERS stars with $\rm{VIPERS_{Zflag}}$ equal to 4, in the same apparent magnitude bin (1 312 objects). 
Cumulatively, our stellar training sample reached 2 232 objects.

Similar to the case of the AGN training sample, we did not divide the stellar training sample in i'-based apparent magnitude bins.
As shown on the representative colour-colour plots for the different magnitude i' bins (Fig.~\ref{STARS}), we did not observe a significant change in the distribution of our stellar sample as a function of apparent luminosity.

\subsection{Oversampling}
\label{oversampling}

Our training sample includes more than 16~000 galaxies, and only 2~232 stars and 498 AGNs. 
Figure~\ref{ALL} shows the representative colour-colour plots for galaxies, AGNs, and stars chosen for the best training sample set.  
Sampling strategies, such as oversampling and undersampling, are popular solutions for tackling the problem of classification because the SVM classifier is sensitive to a high-class imbalance, resulting in a drop in the classification performance \citep[e.g.,][]{tang09, akabani04, raskutti04}.
An unbalanced training set tends to overpredict the majority class for unknown sources \citep{tian11}.

To avoid this effect, we performed an oversampling of the AGN and stellar training sets so that in each considered magnitude bin we had a similar effective number of objects classified as galaxies, AGNs, and stars, respectively. 
In fact, despite our decision not to splits AGN and star classes into magnitude bins, unlike what we did in the case of galaxies, the imbalance between the numbers of representatives in each class remains high. 

\begin{table}
\caption[]{Number (N) of galaxies, AGNs, and stars in our training sample after using the oversampling method.}
\label{OVER}
\begin{tabular}{l|l |l |l|l }
 &19$\leqslant$i'$<$20&20$\leqslant$i'$<$21&21$\leqslant$i'$<$22&22$\leqslant$i'$<$22.5 \\ \hline \hline
N galaxies & 1~884 &5~483&6~778&2~126\\
N AGNs&1~520&4~440&5~440&1~760\\
N stars&2~232&4~440&5~440&2~232\\ \hline
\end{tabular} 
\end{table}

Using a simple oversampling technique, we raised the effective number of AGNs and stars up to $\sim$80\% of the number of galaxies in each magnitude bin considered. 
We therefore added in each magnitude bin a number of artificial objects calculated as
  \begin{equation}
 \rm{\lceil X_{i\_missing}\rceil_{10}= NG_i*0.8-X}
  \end{equation}
  where $\rm{X_{i\_missing}}$ is a number of missing objects (AGNs, stars), and symbol $\lceil \rceil_{10}$ corresponds to rounding the value up to the nearest ten.
  The additional artificial objects were created by shifting the observed magnitudes by an amount drawn from a Gaussian distribution with $\sigma$=0.05. 
We also checked how  the stellar and AGN training samples work if we did not perturb the colours, but instead populated real objects multiple times.  
As might be expected, the results of classifiers were worse than with randomly modified  stars and AGNs. 
This method also allows us to take all possible small residuals differences into account in photometry between the two fields. 
Tab.~\ref{OVER} summarises the numbers of training galaxies in each magnitude-binned set together with the number of  AGNs and stars after oversampling.

\section{Results}
\label{sec_results}

\subsection{Training procedure}
\begin{figure}
 \begin{center}
	\resizebox{0.9\hsize}{!}{\includegraphics{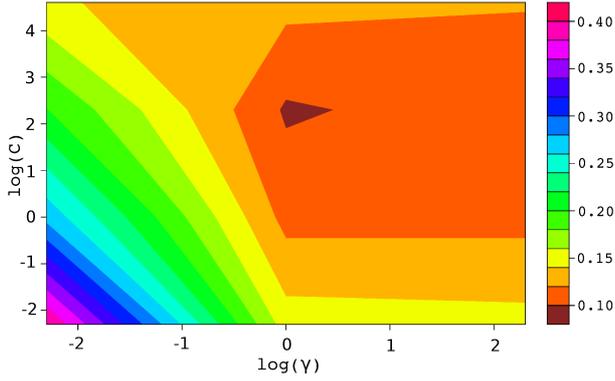}}
	\caption[]{
	The mean misclassification rate as a function of $C$ and $\gamma$ as estimated from the ten-fold cross-validation technique performed for each pair of parameters (see text for more details). 
	The lower the ratio of misclassification, the better the performance of the SVM algorithm. 
	}
\label{Cgamma}
 \end{center}
\end{figure}

To to build a classifier that will be able to separate different classes of objects, it is necessary to tune the \textit{C} and $\gamma$  parameters using the training sample. 
For the best performance, we performed a grid search with values from $\rm{\gamma\in10^{(-3:-1)}}$ and ${C\mbox{ }\rm{\in10^{(0:3)}}}$ using a ten-fold cross-validation technique.  
We first divided the full training sample into ten subsets of equal size and selected nine subsets to train the classification model and test it against the remaining subset (the so-called self-check).  
This test was repeated ten times, with a different subset removed for each training run.  
The classification accuracy was then averaged over the ten runs.  
This process was repeated for each value of the parameters $C$ and $\gamma$. 
In Fig.~\ref{Cgamma} we present a representative plot of the the grid search, done for the apparent magnitude bin 19$\leqslant$i'$<$20. 
The colour of each pointing of the grid codes the mean misclassification rate of all $\gamma$ and \textit{C} values (on a log scale on the X and Y axis, respectively).  
The misclassification rate is defined as (1-total accuracy) for each magnitude bin (see Eq.~\ref{TotalAccuracy} further in the paper): the lower the ratio of misclassification, the better the performance of SVM algorithm. 
We would like to stress that a change in the parameter space (such as adding more parameters describing properties of sources) or a sufficient change in the number of training objects inside one class 
may result in altering the occupancy of training objects and therefore requires recalculatiin the best parameters.

To check the efficiency of our classifiers, we counted the true objects 
(true galaxies - TG, true AGNs - TAGN, and true stars -  TS from the training sample originally classified as  galaxies, AGNs, and stars, respectively) 
and false objects: FG (false galaxy: when a source from the stellar or AGN training sample is classified as a galaxy by the SVM); 
FS (false star: when an object from a galaxy or AGN training sample is classified as a star by the SVM); 
and FAGN (false AGN: when an object from a galaxy or star training sample is classified as an AGN by the SVM). 
We then calculated the accuracy of our classifier based on the formula:

\begin{equation}
\label{Accuracy}
 \rm{Accuracy=\frac{TG+TAGN+TS}{TG+TAGN+TS+FG+FAGN+FS}.}
\end{equation}
After completing the ten-fold cross-validation process we calculated the total accuracy of the SVM classifier, defined as the mean accuracy for all iterations:

\begin{equation}
\label{TotalAccuracy}
 \rm{Total\mbox{ }Accuracy=\frac{\varSigma_{i=1}^{N} Accuracy_{i} }{N},}
\end{equation}
where $N=10$ is the number of validation iterations. 
We performed this check in each magnitude bin considered.

In our work for galaxy/AGN/star classification, we used both a three- and five-dimensional colour space. 
The first one was built using only optical data, corresponding to  (u$^*$-g'), (g'-r), and (r'-i) colours, while the second one included NIR data and thus used two extra colours: (i'-z'), and (z'-$\rm{K_s}$).

\subsection{Optical u$^*$g'r'i' classifier}
\begin{table*}[ht]
\begin{center}
\caption[]{Results of the self-check of the purely optical classifier (u$^*$, g', r', and  i' only). 
Columns corresponds to the true (spectroscopically classified) galaxies, stars, and AGNs. 
Rows correspond to objects classified as galaxies, AGNs, and stars by our classifier. 
Then values in bold correspond to the correctly classified objects (galaxies, AGNs, and stars) in defined i'-based apparent magnitude bins.
Ratios of classified objects are given in percentage.}
\label{SELF_CHECK_WITHOUT_IR_NEW}
\begin{tabular}{l|l|l|l|l|l|l|l|l|l|l|l|l}
               &\multicolumn{3}{|c|}{19$\leqslant$i'$<$20}  &\multicolumn{3}{|c|}{20$\leqslant$i'$<$21} &\multicolumn{3}{|c|}{21$\leqslant$i'$<$22} &\multicolumn{3}{|c}{22$\leqslant$i'$<$22.5}\\ \hline
Total accuracy &\multicolumn{3}{|c|} {85.01\%}     &\multicolumn{3}{|c|}{87.38\%}     &\multicolumn{3}{|c|}{85.09\%}     &\multicolumn{3}{|c}{88.09\%}    \\   \hline \hline
SVM/true       & Galaxy     &AGN &Star& Galaxy     &AGN &Star& Galaxy     &AGN &Star& Galaxy     &AGN &Star\\ \hline
Number of sources& 1~884&1~520&2~232&5~483&4~440&4~440&6~778&5~440&5~440&2~126&1~760&2~232\\ \hline
Galaxy &\textbf{88.82}  &15.70  &10.98   &\textbf{92.10} &6.23   &15.06  &\textbf{88.39} &15.50  &10.01  &\textbf{93.18}  &17.47 &3.00\\
AGN    &4.45  &\textbf{69.45}&   10.23   &3.28   &\textbf{90.88}  &4.48  &4.04  &\textbf{81.54}  &3.81  &4.37  &\textbf{79.06}  &3.28\\
Star   &6.73  &14.85    &\textbf{78.79}  &4.62   &2.89  &\textbf{80.46}  &7.57   &2.96  &\textbf{86.19}  &2.46  &3.47  &\textbf{93.72}\\ \hline
\end{tabular} 
\end{center}
\end{table*}

\begin{table*}
\begin{center}
\caption[]{Results of the self-check of the classifier with the near-infrared data (u$^*$, g', r',  i', z', and $\rm{K_s}$). 
Columns correspond to the true (spectroscopically classified) galaxies, stars, and AGNs. 
Rows correspond to objects classified as galaxies, AGNs, and stars by our classifier.
The values marked in bold are correctly classified objects (galaxies, AGNs, and stars) in defined i'-based apparent magnitude bins.
Ratio of classified objects are given in percentage.}
\label{SELF_CHECK_WITH_IR}
\begin{tabular}{l|l|l|l|l|l|l|l|l|l|l|l|l}
               &\multicolumn{3}{|c|}{19$\leqslant$i'$<$20}  &\multicolumn{3}{|c|}{20$\leqslant$i'$<$21} &\multicolumn{3}{|c|}{21$\leqslant$i'$<$22} &\multicolumn{3}{|c}{22$\leqslant$i'$<$22.5}\\ \hline
Total accuracy &\multicolumn{3}{|c|} {95.47\%}     &\multicolumn{3}{|c|}{95.83\%}     &\multicolumn{3}{|c|}{94.28\%}     &\multicolumn{3}{|c}{94.58\%}    \\   \hline \hline
SVM/true       & Galaxy     &AGN &Star& Galaxy     &AGN &Star& Galaxy     &AGN &Star& Galaxy     &AGN &Star\\ \hline
Number of sources& 1~884&1~520&2~232&5~483&4~440&4~440&6~778&5~440&5~440&2~126&1~760&2~232\\ \hline
Galaxy &\textbf{96.28} &2.90 &1.27  &\textbf{97.61} &1.95 &0.44  &\textbf{97.11} &5.00 &2.10   &\textbf{96.10} &6.09 &1.57\\
AGN    &2.44 &\textbf{95.91} &1.70  &1.95 &\textbf{96.34} &0.80  &2.52 &\textbf{94.83} &0.77   &3.38 &\textbf{92.94} &1.30\\
Star   &1.28 &1.19 &\textbf{96.37}  &0.44 &0.27 &\textbf{97.25}  &0.37 &0.17 &\textbf{97.13}   &0.52 &0.97  &\textbf{97.13}\\
\end{tabular}
\end{center}
\end{table*}

We constructed colour-colour training samples without near-infrared data, based only on the optical  u$^*$, g', r', and  i' filter bands (a three-dimensional hyperspace).
We found that the Total Accuracy, as well as the number of  correctly classified objects for this approach, depend on the apparent magnitude of objects. 
Averaging over all magnitude bins (19$\leqslant$i'$<$22.5), once we average results by the number of objects in each bin, the mean Total Accuracy for the optical classifier is equal to 86.39\%.

The results of the self-check of our classifier are shown in Table~\ref{SELF_CHECK_WITHOUT_IR_NEW}, showing that only in a few percent of the cases (less than 11\% in all magnitude bins), galaxies are classified as a star or as an AGN. 
The most frequent misclassifications occur in the 19$\leqslant$i'$<$20 bin, in which galaxies are correctly classified at the level of 88.82\%, AGNs - 69.45\%,  and stars at the level of 78.79\%.
The misclassifications between stars and galaxies are noticeable in the first three bins. 
For 20$\leqslant$i'$<$21 and 21$\leqslant$i'$<$22 bins, more than 10\% of spectroscopically classified stars are classified by the SVM as false galaxies (15.06\% and 10.01\%, respectively). 
In the same bins, AGNs are mis-classified as galaxies at the high levels of 6.23\% and 15.50\%, respectively.

The misclassification of galaxies and AGNs happens mainly in the bins where the percentage of oversampled objects increases. 
The reason may be related either to our oversampling method or to the lower accuracy of photometry for the fainter sources, as well as to the intrinsic properties of classified sources in these bins. 
We stress that for the SVM method the 100\% level of self-check is not desirable since it may indicate over-fitting.  
The boundaries between different classes of objects defined by the training sample may become too rigid and artificially complex, not allowing for effective classification  of real sources.   
Nevertheless, it seems that the present, very basic classifier, which was created on the basis similar to the standard colour-colour approach, works well for our training sample. 
 
\begin{table}[h]
\caption[]{Test of SVM optical classifier  on the galaxies with $\rm{VIPERS_{Zflag}}$ equal to 3. 
In the first row we show the percentage of correctly classified galaxies. 
Second and third rows show the percentage of miss-classified galaxies: when a true galaxy is classified by SVM as an AGN or a star, respectively.}
\label{GAL3_withoutIR_NEW}
\begin{tabular}{l|l|l|l|l}
 & 19$\leqslant$i'$<$20 & 20$\leqslant$i'$<$21 & 21$\leqslant$i'$<$22 & 22$\leqslant$i'$<$22.5 \\ \hline \hline
Galaxies     &90.97  &91.41 &85.38 &88.82\\
False AGNs   &2.76   &2.81  &3.06  &4.45\\
False stars  &6.27   &5.78  &11.56 &6.73\\ \hline
\end{tabular} 
\end{table}

We next apply our trained classifier to VIPERS galaxies with redshift quality flag $\rm{VIPERS_{Zflag}}$ = 3, corresponding to a confidence of the redshift measurements - and correspondingly of correct identification as a galaxy - of $>99\%$ (hereafter $\rm{GAL_{3}}$).  
Table~\ref{GAL3_withoutIR_NEW}  shows that $\rm{GAL_{3}}$ are correctly classified at a level higher than 85\%  with a percentage of misclassification that is almost constant at a level of 15\% maximum. 
The strong contamination by false stars is visible for objects fainter than i'=21 mag. 
It is reassuring that this trend is similar to the self-check results (Table~\ref{SELF_CHECK_WITHOUT_IR_NEW}) demonstrating that the training sample is representative of the data.
In the fainter magnitude bins, the photometric errors increase such that the optical u$^*$, g', r', and i' fluxes are not as efficient in distinguishing galaxies and stars. 

\subsection{Optical+NIR (u$^*$g'r'i'z'$\rm{K_s}$) classifier}

We enlarged the parameter space by adding the  NIR colours (z' and $\rm{K_s}$) to our classifier (a five-dimensional hyperspace). 
We performed the same tests as for the optical classifier (self-check, and test on VIPERS $\rm{GAL_{3}}$).  

Our training sample, composed of exactly the same sources as the optical classifier, but with NIR measurements, allows us to train a new optical + NIR classifier.  
The  mean Total Accuracy for this classifier is equal to 94.29\%, i.e. higher than the pure optical one.
Total accuracy for particular magnitude bins stays on the similar level $\sim$ 95\% for the whole i'-apparent magnitude binned sample. 
The constancy of the new classifier for objects fainter than 20 mag in i' band is very promising for the next tests and final classification of VIPERS objects. 

Table~\ref{SELF_CHECK_WITH_IR} shows the self-check for the u$^*$, g', r', i', z' and $\rm{K_s}$ space classifier.
When we average over all magnitude bins, galaxies are correctly classified in $\sim$ 97.03\%, AGNs in 95.13\%,  and stars in  97.05\% of the cases. 
All these numbers are significantly higher than those for a purely optical classifier. 
In the case of AGNs, the difference between correctly classified sources for optical and optical+NIR classifiers is equal to 26.46\%, 5.46\%, 13.30\%, and 13.88\% for 19$\leqslant$i'$<$20, 20$\leqslant$i'$<$21, 21$\leqslant$i'$<$22, and 22$\leqslant$i'$<$22.5 apparent magnitude bins, respectively. 
Stars are correctly classified at a higher level than AGNs, with a difference between optical and optical+NIR classifiers equal to 17.58\%, 16.79\%, 10.94\%, and 3.41\%  for the same magnitude bins.  

\label{ClassifierWithIR}
\begin{table}
\caption[]{Test of SVM classifier with near-infrared data on the galaxies with $\rm{VIPERS_{Zflag}}$ equal to 3. 
The first row represents  the percentage of correctly classified galaxies. 
Second and third rows show the percentage of mis-classified galaxies: when a true galaxy is classified by SVM as an AGN or a star, respectively.}
\label{GAL3_withIR_NEW}
\begin{tabular}{l|l|l|l|l }
 & 19$\leqslant$i'$<$20&20$\leqslant$i'$<$21&21$\leqslant$i'$<$22&22$\leqslant$i'$<$22.5 \\ \hline \hline
Galaxies     &95.38 &95.17 &93.09 &92.72\\
False AGNs   &2.42  &2.72  &4.30  &5.29\\
False stars  &2.20  &2.11  &2.61  &1.99\\
\end{tabular} 
\end{table}

\begin{figure*}
\begin{minipage}[t]{0.49\textwidth}
 	\includegraphics[width=1\textwidth]{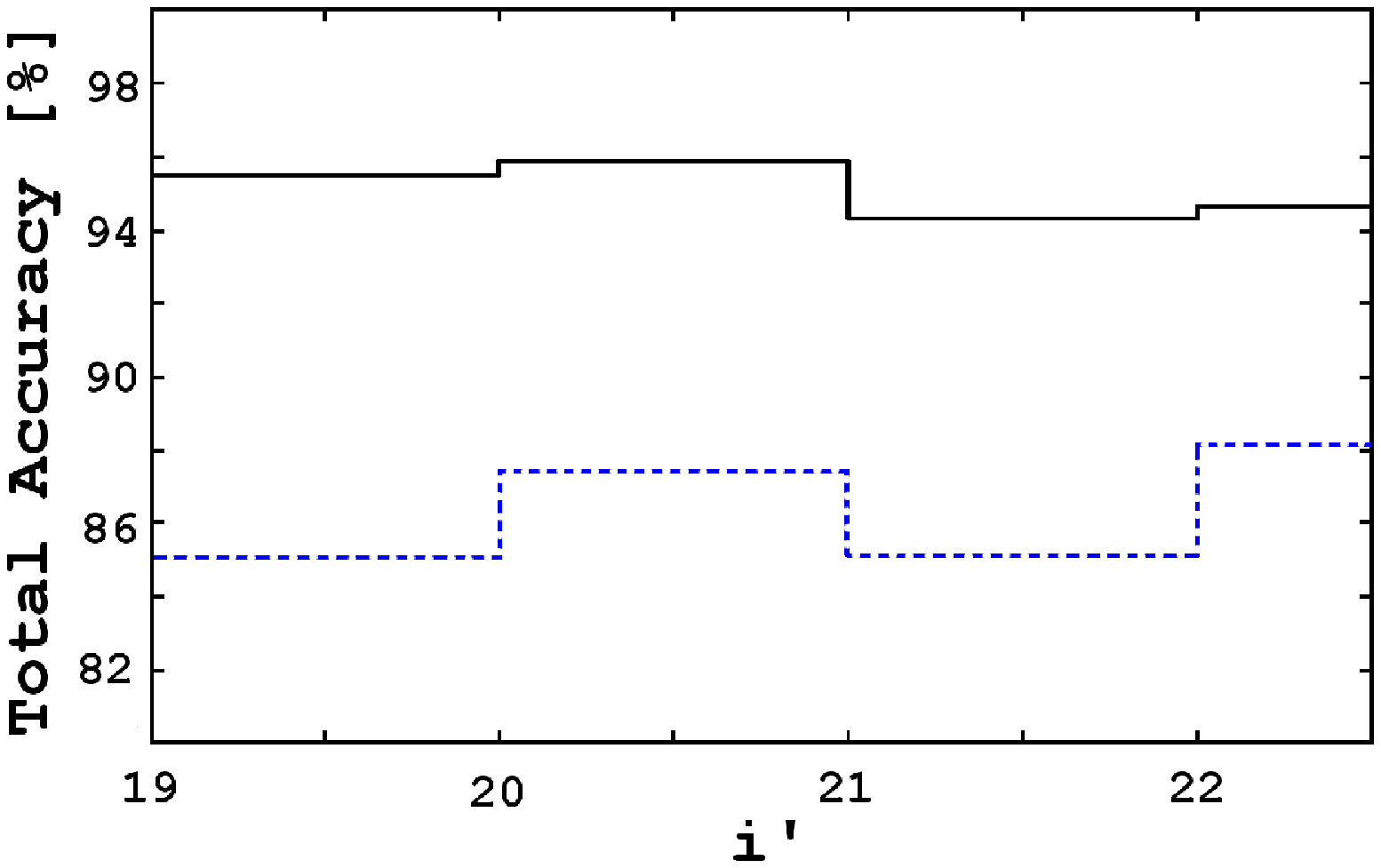}
	\caption[]{Total accuracy for optical and optical+NIR classifiers (see Tables~\ref{SELF_CHECK_WITHOUT_IR_NEW}~and~\ref{SELF_CHECK_WITH_IR}).
	Results for the optical classifier based on the  u$^*$, g', r', and i' filter are marked as a dotted line.
	Solid line corresponds to the total accuracy of the optical+NIR classifier.}
	\label{TA1}
  \end{minipage}\hfill 
  \begin{minipage}[t]{0.49\textwidth} 
  \includegraphics[width=1\textwidth]{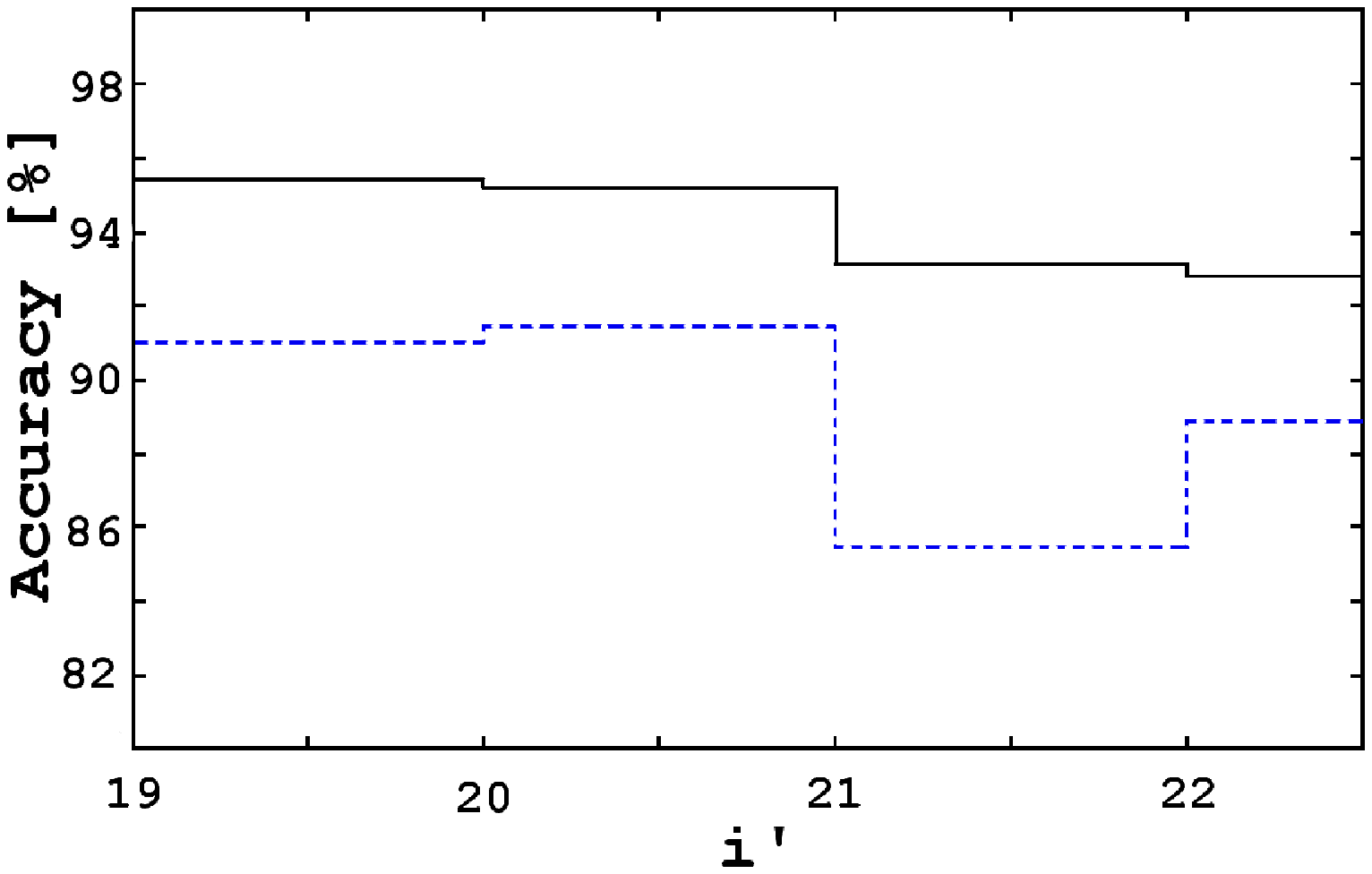}
	\caption[]{The accuracy of optical and optical+NIR classifiers for VIPERS galaxies with $\rm{VIPERS_{Zflag}}$ equal to 3 
($\rm{GAL_{3}}$).
	Results for classifier based on the  u$^*$, g', r', and i' filters only are marked as a dotted line.
	Solid line corresponds to the classifier with the NIR data (u$^*$, g', r', i', z', and K$_s$).
	}
	\label{GAL3_fig1}
\end{minipage} 
\end{figure*}

Applying this classifier to VIPERS galaxies with  $\rm{VIPERS_{Zflag}}$ equal to 3 
($\rm{GAL_{3}}$, Table~\ref{GAL3_withIR_NEW}) shows that galaxies are correctly classified at the very high level of 93.60\% (we average results by the number of objects in each bin).
Incorrect galaxy classifications, false AGNs and false stars, are very rare and do not exceed  2.65\% for stars and 5.30\% for AGNs.

We can observe the trend for galaxies to have an increased risk of being misclassified as AGNs in the faintest magnitude bins. 
One possible explanation for this behaviour is the decrease in the quality of the photometry for the less luminous sources, which have a lower signal-to-noise ratio. 
On the other hand, the limiting magnitude of CFHTLS is much deeper than the VIPERS one, and photometry should still be fairly good down to mag i' 22.5. 
Another explanation could be that some of these galaxies are hosting faint AGNs that were not recognised  during the visual verification and validation of the measured redshift, since with the decreasing luminosity the host galaxy becomes dimmer and the AGN component becomes more significant.
This possibility will be examined further in future works.

\subsection{Comparison of the classifiers}

In Fig.~\ref{TA1} we compare the  total accuracy for the optical and optical+NIR classifiers.  
However, on average the classifier based on the u$^*$, g', r', i', z', and $\rm{K_s}$  bands is 7.90\% better then the classifier trained without z' and $\rm{K_s}$  data. 
Moreover, the total accuracy of the optical+NIR classifier decreases very weakly with the apparent magnitude, while a strong variation from bin-to-bin is visible for the purely optical classifier.
etween the first and the second apparent magnitude bin the difference between their total accuracy rises from  6.49\% to 10.46\% from the fainter to the brighter bins. 

The preponderance of the classifier constructed  with the NIR data is confirmed by  the efficiency of correctly classifying  of galaxies with $\rm{VIPERS_{Zflag}}$ equal to 3 ($\rm{GAL_{3}}$).
Figure~\ref{GAL3_fig1} shows the comparison of accuracy of both classifiers (with and without near-infrared data) for the $\rm{GAL_{3}}$ sample. 
For the fainter objects (21$\leqslant$i'$<$22), the efficiency decreases rapidly for the classifier trained without z' and $\rm{K_s}$ bands, and much smoother for the more sophisticated classifier trained with infrared features.

We conclude that including NIR data to train the SVM algorithm significantly improves the efficiency of the galaxy/AGN/star classifier.
It is evident that NIR features are very important for building an effective classifier for basic astronomical classification of these three classes of sources. 
Based on the above tests, we decided to choose the classifier based on the u$^*$, g', r', i', z', and $\rm{K_s}$ bands to be used in our next analysis.

\begin{figure}[t]
 \begin{center}
	\resizebox{0.9\hsize}{!}{\includegraphics{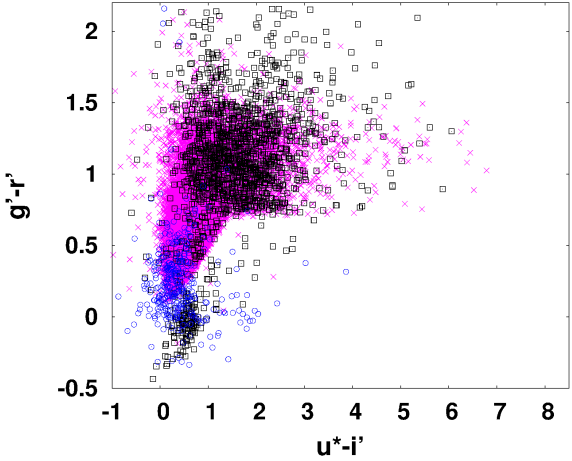}}
	\caption[]{The representative colour-colour plot for all objects used for a consistency check for VIPERS objects with redshift confirmation levels $>$ 99\%, with i' apparent magnitude between 19 and 22.5. 
	Pink \texttt{x}-s represents galaxies with $\rm{VIPERS_{Zflag}}$=3. 
	Open blue circles correspond to AGN sample with redshift confirmation level equal to or higher than 99\% ($\rm{VIPERS_{Zflag}}$ equal to 13 and 14). 
	Open black squares correspond to stellar sample with $\rm{VIPERS_{Zflag}}$ equal to 3 and 4. 
	}
\label{TESTVIPERS3}
 \end{center}
\end{figure}

\section{Consistency checks on VIPERS data}
\label{test_VIPERS_data}

\subsection{VIPERS objects with redshift confirmation level of~$\geqq$~99\%}
We now apply the optical+NIR classifier  only to VIPERS data:
\begin{itemize}
 \item galaxy sample - all ($\rm{GAL_{3}}$) galaxies in i'-apparent magnitude range between 19 and 22.5 mag, with the total number of sources equal to 13 539,
 \item AGN sample - all AGNs detected by VIPERS, with redshift confirmation level equal to or higher than 99\%, and with i' apparent magnitude between 19 and 22.5 (367 objects). 
 All of these AGNs were used to build the training sample (see Sect.~\ref{AGNsTS}) which means that our classifier should know their position in our five-dimensional space of parameters,
 This is not as worrisome as it may look thanks to the high oversampling needed for AGN sample (more than 200\% for the brightest and the faintest apparent magnitude bins, and almost 800\% for 20$\leqslant$i'$<$21 and 21$\leqslant$i'$<$22 for i'-apparent magnitude bins) that significantly erases the possibly  peculiar characteristics of the 367 AGN chosen for the training sample. 
 \item stellar sample - all spectroscopically detected stars, with confirmation level of $>$ 99\% ($\rm{VIPERS_{Zflag}}$ equal to 3 and 4), and i' apparent magnitude between 19 and 22.5 (1 729 stars). 
 All stars with $\rm{VIPERS_{Zflag}}$=4 were used as a part of stellar training sample. 
\end{itemize}
Figure~\ref{TESTVIPERS3} shows the representative colour-colour plot for $\rm{GAL_{3}}$, AGNs with $\rm{VIPERS_{Zflag}}$ equal to 13 and 14, and stars with $\rm{VIPERS_{Zflag}}$ equal to 3 and 4, chosen for the consistently check.

\begin{table*}[ht]
\begin{center}
\caption[]{Results of the test of the optical+NIR classifier for $\rm{GAL_{3}}$, and  AGNs and stars with redshifts measurements on a  confirmation level $\geq$ to 99\%.
Values marked in bold correspond to the correctly classified objects (galaxies, AGNs, and stars) in i'-based apparent magnitude bins.
The ratio of the classified objects is given in percentage.}
\label{TEST_GAl_AGN_STARS_flaga3}
\begin{tabular}{l|l|l|l|l|l|l|l|l|l|l|l|l}
               &\multicolumn{3}{|c|}{19$\leqslant$i'$<$20}  &\multicolumn{3}{|c|}{20$\leqslant$i'$<$21} &\multicolumn{3}{|c|}{21$\leqslant$i'$<$22} &\multicolumn{3}{|c}{22$\leqslant$i'$<$22.5}\\ \hline
SVM/true       & Galaxy     &AGN &Star& Galaxy     &AGN &Star& Galaxy     &AGN &Star& Galaxy     &AGN &Star\\ \hline
Number of sources& 445&69&337&3~271&1340&428&7~667&127&701&2~156&37&263\\ \hline
Galaxy  &\textbf{95.38} &12.52 &4.17   &\textbf{95.17} &7.37 &3.27   &\textbf{93.09} &10.46 &3.42   &\textbf{92.72} &14.90 &9.09     \\
AGN     &2.42 &\textbf{77.34} &3.70   &2.72 &\textbf{82.08} &3.27   &4.30 &\textbf{82.75}  &1.43   &5.29 &\textbf{75.54} &6.44        \\
Star    &2.20 &10.14 &\textbf{92.13}   &2.11 &10.55 &\textbf{93.46}  &2.61 &6.79 &\textbf{95.15}    &1.99 &9.56 &\textbf{84.47} \\ \hline
\end{tabular}
\end{center}
\end{table*}

For this test, all three classes of sources were  divided into four i'-apparent magnitude bins (19$\leqslant$i'$<$20, 20$\leqslant$i'$<$21, 21$\leqslant$i'$<$22, and 22$\leqslant$i'$<$22.5), the same as  used in the training sample. 
Then, we applied our optical+NIR classifier to this data. 
Tab.~\ref{TEST_GAl_AGN_STARS_flaga3} shows the results of the automatic classification. 

The mean accuracy for galaxies, averaged over the mean number of objects  in each apparent magnitude bin, equals 93.60\%.
This result for galaxy classification displays only  aslightly lower level of efficiency ($\sim$1.50\%) than the galaxy classification obtained during the self-check of the classifier (see Sect.~\ref{ClassifierWithIR}). 
It means that the hyperspace of galaxy parameters used for the training sample is well defined. 

The result of AGN classification is worse than the one obtained during the self-check but still satisfactory.
After averaging over all magnitude bins, AGNs are correctly classified at a level equal to 81.80\% with a significant decrease  with i' apparent magnitude between 21 and 22 mag. 
Stars are correctly classified at the high mean level of 92.52\% with a significant drop for the 22$\leqslant$i'$<$22.5 apparent magnitude bin (84.47\%). 
The performance of the classifier in the case of AGNs may look relatively poor. 
However, as already mentioned, we should remember that the VIPERS selection allows AGNs preclassified as galaxies or stars based on their colour properties. 
Keeping this in mind, we should instead feel satisfied that a high fraction of these AGNs can be separated into a different section of the five-dimensional hyperspace from galaxies and stars, when using an AGN training sample that only consists of 498 objects.   

We did not find any crucial misclassifications for the galaxy sample. 
The galaxies are classified correctly on a very high level.
For the AGN sample, the contamination of true AGNs classified as  galaxies (8.17\%, 7.37\%, 10.46\%, 14.90\% for the 19$\leqslant$i'$<$20, 20$\leqslant$i'$<$21, 21$\leqslant$i'$<$22, and 22$\leqslant$i'$<$22.5 bins, respectively) and stars (8.96\%, 10.55\%, 6.79\%, 9.56\% for the 19$\leqslant$i'$<$20, 20$\leqslant$i'$<$21, 21$\leqslant$i'$<$22, and 22$\leqslant$i'$<$22.5 bins, respectively) is significant. 
For the stellar sample, the classifier  misclassified true stars as  galaxies more often than AGNs.
In the future development of this classifier, we will include the morphological information, as well as emission/absorption lines, which should improve the algorithm and increase the percentage of correctly classified sources as well. 
Including the morphological information will allow us to construct a classifier that could be applied to purely photometric surveys, similar to the one
presented in this paper. 
Adding spectroscopic information to the parameter space would restrict the use of the classifier, but it would allow for more precise classification schemes.

\subsection{VIPERS objects with redshift confirmation level lower than 99\%}

We performed a classification for VIPERS objects with confirmation levels lower than 99\%.
In particular, we used galaxies, AGNs, and stars from the VIPERS database, with the quality of the measured  redshift, $\rm{VIPERS_{Zflag}}$,  equal to two and one. 
$\rm{VIPERS_{Zflag}}$ equals two means that the measured redshift is fairly secure, with a confidence level  $\geqslant$ 95\%. 
Objects with $\rm{VIPERS_{Zflag}}$ equal to one are more tentative, and their redshift measurement was based on weak spectral features and/or continuum shape.
For these objects there is a $\sim$∼ 50\% probability that the redshift could be wrong.
A more detailed description of the $\rm{VIPERS_{Zflag}}$ and quality of measured redshifts can be found in \citet{guzzo13}, and \citet{garilli12}.

\begin{table*}[ht]
\begin{center}
\caption[]{Results of the optical + NIR classifier for galaxies, AGNs, and stars with redshifts measurements on a  confirmation level equal to 95\% ($\rm{VIPERS_{Zflag}}$ = 2).
Objects are not related to the training sample.
Values marked in bold correspond to the correctly classified objects (galaxies, AGNs, and stars) in i'-based apparent magnitude bins.
The ratio of the classified objects is given in percentage.
}
\label{dwojki}
\begin{tabular}{l|l|l|l|l|l|l|l|l|l|l|l|l}
               &\multicolumn{3}{|c|}{19$\leqslant$i'$<$20}  &\multicolumn{3}{|c|}{20$\leqslant$i'$<$21} &\multicolumn{3}{|c|}{21$\leqslant$i'$<$22} &\multicolumn{3}{|c}{22$\leqslant$i'$<$22.5}\\ \hline
SVM/true       & Galaxy     &AGN &Star& Galaxy     &AGN           &Star& Galaxy     &AGN &Star& Galaxy     &AGN &Star\\ \hline
Number of sources & 8 & 33           &10           &945           &75             &27            &5~757          &80            &145           &6~226          &48            &159\\ \hline
Galaxy &\textbf{84.15} &6.25  &30.00    &\textbf{94.18}&12.00 &22.22  &\textbf{92.81} &20.25 &29.65  &\textbf{58.62} &4.17  &17.61     \\
AGN    & 9.75 &\textbf{93.75} &40.00    &3.49 &\textbf{88.00} &29.63      &4.41 &\textbf{77.22} &11.03   &20.56 &\textbf{93.75} &18.87        \\
Star   & 6.10 & 0.00 &\textbf{30.00}    &2.33 &0.00 &\textbf{48.15}    &2.78 &2.53 &\textbf{59.32}    &20.82 &2.08 &\textbf{63.52} \\ \hline
\end{tabular}
\end{center}
\end{table*}

\begin{table*}[ht]
\begin{center}
\caption[]{The results of the optical + NIR classifier for galaxies, AGNs, and stars with redshifts measurements on a  confirmation level equal to 50\% ($\rm{VIPERS_{Zflag}}$ equals to 1).
Objects are not connected with the training sample.
Values marked in bold correspond to the correctly classified objects (galaxies, AGNs, and stars) in i'-based apparent magnitude bins. 
The ratio of the classified objects is given in percentage.
}
\label{jedynki}
\begin{tabular}{l|l|l|l|l|l|l|l|l|l|l|l|l}
               &\multicolumn{3}{|c|}{19$\leqslant$i'$<$20}  &\multicolumn{3}{|c|}{20$\leqslant$i'$<$21} &\multicolumn{3}{|c|}{21$\leqslant$i'$<$22} &\multicolumn{3}{|c}{22$\leqslant$i'$<$22.5}\\ \hline
SVM/true       & Galaxy     &AGN &Star& Galaxy     &AGN &Star& Galaxy     &AGN &Star& Galaxy     &AGN &Star\\ \hline
Number of sources &35 &8 &4 &355 &13 &24 &2~833 &35 &81 &3~157 &30 &139 \\ \hline
Galaxy &\textbf{85.71} &50.00  &25.00    &\textbf{92.68} &23.08 &37.50  &\textbf{88.14} &20.00 &60.56  &\textbf{43.09} &23.33 &25.90       \\
AGN    &11.43 &\textbf{50.00} &75.00    &3.94 &\textbf{76.92} &37.50   &7.12 &\textbf{68.57} &9.86    &31.03 &\textbf{66.67} &31.65        \\
Star   &2.86  &0.00 &\textbf{0.00}      &3.38 & 0.00 &\textbf{25.00}   &4.74 &11.43 &\textbf{29.58}    &25.88 &10.00 &\textbf{42.45} \\ \hline
\end{tabular}
\end{center}
\end{table*}

Results of SVM classification of objects with $\rm{VIPERS_{Zflag}}$ equal to two (Table~\ref{dwojki}) and one (Table~\ref{jedynki}) show very good conformity to the previously user supervised estimations. 
Galaxies are classified with agreement to redshift measurements on the mean level of  76.45\%  for $\rm{VIPERS_{Zflag}}=2$ and 66.08\% for  $\rm{VIPERS_{Zflag}}=1$.

The ongoing scientific analysis of galaxy evolution and clustering is mainly based on objects that have  secure redshift measurements ($\rm{VIPERS_{Zflag}} \geq$ 2, depending on the topic). 
With the SVM classification, we can reconfirm the identify of galaxies with the lower quality flags and thus increase the number of galaxies that could be used for more detailed analysis. 
This may apply to 4~735 galaxies, 58 AGNs, and 86 stars with $\rm{VIPERS_{Zflag}}$ equal to one\footnote{These numbers were calculated as a sum of galaxies, AGNs, and stars which were  classified to the same class of objects during redshift validation and by an optical+NIR classifier; marked in bold in Table~\ref{jedynki}.}.  
This method may also reconfirm the class of 9~952 galaxies, 177 AGNs,  and 160 stars  with  $\rm{VIPERS_{Zflag}}=2$ classified as galaxies, AGNs, and stars by checking the results twice by different observers, and by our classifier. 

One may argue that the $\rm{VIPERS_{Zflag}}$ is related to the redshift value, not to the identification of the galaxy itself. 
However, it should be noted that a majority of sources with low $\rm{VIPERS_{Zflag}}$ are absorption line systems with noisy, low signal-to-noise spectra. 
Galaxies with such spectra can be particularly easily misclassified as stars during the spectroscopic measurement process, either automatic or human-supervised. 
For instance, typical features of an elliptical galaxy at z$\sim$1, around the Balmer break, can be confused with characteristic features of an M-type star. 
To confirm the redshift measurements or flag validation, it is possible to use SED templates for the photometric redshift estimation, and to compare spectroscopic redshifts with photometric ones, but SED-fitting for sources with poor photometry can be degenerate possibly leading to biased results. 
In such cases, an independent confirmation that the position of an object in the five-dimensional colour space is actually typical of galaxy, and actually increases also probability that its redshift has been assigned correctly.

The number of stars and AGNs in the sample of objects with $\rm{VIPERS_{Zflag}}$ equal to one and two is very low (a few objects in the brightest apparent luminosity bin, and a few dozen for objects with i' magnitude lower than 21 mag). 
This fact results from an initial star/galaxy separation performed by VIPERS. 
In the VIPERS database, the stars, which remained after the colour-colour preselection, are not typical, and they occupy a similar area to galaxies on the colour-colour plots.
Then, that we can reconfirm the identity of a significant fraction of them can already be regarded as a success. 
The spectra of stars with $\rm{VIPERS_{Zflag}}=1$ or 2, which were classified as galaxies by our classifier, will have to be re-examined since some of them might also be genuine galaxies. 

We should consider that the VIPERS galaxy sample is not pure and includes some AGN types such as those with narrow-line features, even for objects with $\rm{VIPERS_{Zflag}}>3$. 
During the standard redshift measurements process only the broad line AGNs are being recognised and flagged. 
This implies that that our galaxy training sample alco contains, in addition to a pure sample of normal galaxies, specific types of AGNs, otherwise difficult to recognise in VIPERS spectra during a standard redshift measurements process.  
The VIPERS hunt for the specific types of AGNs  lurking within the heap of collected sources is still going on, so we are forced to work with the data composed of both galaxies and AGNs, at least for now.
The contamination of galaxies and AGNs is most prominent for the faintest bin (22$\leqslant$i'$<$22.5), where more than 20\% (30\%) of objects classified as galaxies with $\rm{VIPERS_{Zflag}}$ equal to two (one) are identified as AGNs by an optical+NIR SVM  classifier. 

We look forward to using SVM methods to add more information on spectral lines and source morphologies as a very promising tool to improve classification for fainter sources and to refine further classes of objects that the software can discriminate.

\section{Comparison with combined spectral energy distribution fitting and geometric method}
\label{comaprison}
As a test of efficiency of the SVM VIPERS classifier, we  compared our algorithm with the star/galaxy separtion of the VVDS data performed by Coupon  \citep{coupon09,guzzo13}.  
We computed the incompleteness of our galaxy selection as the ratio of true galaxies/AGNs lost after SVM classification, and we defined contamination as a number of stars mis-classified by SVM as galaxies/AGNs.

\cite{coupon09} base their star/galaxy classification on the most secure spectroscopic sample from the VVDS F02 and VVDS F22 fields. 
The method adopted for the star/galaxy separation was a combination of a geometric method for objects brighter than i'=21 mag (half-light radius parameter, r$\rm{_{h}}$, defined as the radius containing half of 
the object's flux,  which was provided by the CFHTLS database), and a combination of geometric and photometric methods for objects fainter than i'=21 mag, fitting u$^*$, g', r', i', and z' bands by a set of the SED templates with the Le Phare photometric redshift code \citep{arnouts99, ilbert06}. 
For a detailed description of this method we refer the reader to \cite{guzzo13}.

\subsection{Sample selection}
We performed  a star/non-star (where non-star for our classifier means galaxy+AGN) selection using the VVDS Deep F02 survey matched with the CFHTLS photometric catalogue (T0005 data release).
We decided to perform star/galaxy classification in the VVDS Deep F02 only, because the stellar training sample used for our classifier was built from stars from the VVDS Wide F22 field. 
Only a part of AGNs from the VIPERS survey was used to train our algorithm.
As a result, our stellar/galaxy separation in this field would be treated preferentially, which could bias the results.

For our test we selected objects with the most secure VVDS flags ($\rm{VVDS_{Zflag}}$ equal to 3, and 4). 
In the next step we selected objects using the same colour/redshift criteria as applied to the VIPERS survey:
\begin{equation}
 \rm{(r-i)>0.5*(u-g)\mbox{ } or \mbox{ }(r-i)>0.7.}
\end{equation}
For the more detailed description and the origin of this colour-based selection, we refer the reader to \citet{guzzo13}.
Then, we divided our sample into two subsamples: 
\begin{enumerate}
 \item non-stars with spectroscopic redshift $\geqslant$ 0.01, and
 \item stars with spectroscopic redshifts $\leqslant$ 0.01,
\end{enumerate}
and then into i' apparent magnitude-binned samples. 
We stress that the ratio of AGNs is not known within the galaxy sample in this case.

\subsection{Method}
The SVM opt+NIR galaxy/AGN/star classifier was applied to this data set. 
We computed the incompleteness and contamination for selected non-stars (galaxies and AGNs). 
Since we did not use geometric selection based on the r$\rm{_h}$, we decided to perform the comparison with the \cite{coupon09} method for  only the fainter bins (21$\leqslant$i'$<$22, and 22$\leqslant$i'$<$22.5), 
where star/galaxy separation was performed based not only on the geometrical properties of sources, but also by fitting SEDs.  
We defined the incompleteness (\texttt{INC}) and contamination (\texttt{CON}) ratio, following \citet{coupon09} and \citet{guzzo13}, as
\begin{equation}
\label{INC}
 \rm{INC=\frac{NG_{true}-NG_{SVM good}}{NG_{true}}},
\end{equation}
and
\begin{equation}
 \rm{CON=\frac{NG_{SVM bad}}{NG_{SVM estimated}}},
\end{equation}
 where
\begin{itemize}
 \item $\rm{NG_{true}}$ is a total number of spectroscopically classified 
 non-stars in the VVDS Deep F02 field with the most secure redshift quality flag ($\rm{VVDS_{Zflag}}$ 3 and 4), 
 \item $\rm{NG_{SVM good}}$ - a number of real non-stars (galaxies and AGNs) classified by SVM algorithm as non-stellar objects,
 \item $\rm{NG_{SVM bad}}$ - a number of galaxies/AGNs mis-classified by our classifier as stars,  and 
 \item $\rm{NG_{SVM estimated}}$ is a total number of objects classified by SVM as a galaxies or AGNs.
\end{itemize}
 
\begin{table}[ht]
\begin{center}
\caption[]{(Galaxy+AGN)/star selection results: the incompleteness and contamination of the VIPERS galaxy sample (VVDS-Deep F02 field) expected from the star-galaxy separation process adopted in VIPERS, and from the SVM opt+NIR classifier.}
\label{INC_CON}
 \begin{tabular}{l|l|l|l|l}
Apparent magnitude & \texttt{INC} & \texttt{CON} & \texttt{INC} & \texttt{CON} \\ \hline \hline
 &\multicolumn{2}{c}{\citet{guzzo13}}& \multicolumn{2}{|c}{SVM}\\ \hline
21$\leqslant$i'$<$22   & \multirow{2}{*}{2.07\%} & \multirow{2}{*}{0.87\%} & 2.13\% & 2.39\% \\
22$\leqslant$i'$<$22.5 &  &  & 2.05\% & 2.00\% \\ \hline
 \end{tabular}
 \end{center}
\end{table}

\subsection{Results}
Table~\ref{INC_CON} shows the results of incompleteness and contamination of two  classifiers applied to the VVDS-Deep F02 sample: one based on the SEDs and geometrical properties of sources \citep{guzzo13} and the other based on the SVM method. 
Presented values are the expected once for incompleteness and contamination of the star/galaxy separation in the W1 VIPERS field.

The incompleteness for both methods is similar (2.07\%  from \citealp{guzzo13} vs 2.13\% and 2.05 for SVM). 
The stellar contamination in the SVM method is slightly higher (0.87\% vs 2.39\% and 2.00\% for SVM algorithm). 
Comparing these results with the results of the self-check of our classifier (see Table~\ref{SELF_CHECK_WITH_IR}), we conclude that the high stellar contamination for these bins might be  related to misclassifications between AGNs and stars\footnote{More than 5\% of real AGNs were classified by our algorithm as stars.}.
Unfortunately, this conclusion cannot be compared directly with the \cite{coupon09} method because theirs does not classify AGNs. 

We checked the real stellar contamination in the W1 field after VIPERS spectroscopic measurements. 
In total, 264 from the 23~360 objects in the 21$\leqslant$i'$<$22.5 apparent magnitude bin preclassified as galaxies using the SEDs+r$\rm{_h}$ method from the PDR-1 catalogue were spectroscopically classified as a stars with $\rm{VIPERS_{Zflag}}\geqslant$1.  
It means that the real contamination on W1 VIPERS filed for object fainter than i'$\geqslant$21 mag is equal to
\begin{equation}
 \rm{CON_{VIPERS\_W1\_i'\geqslant21}=\frac{264}{23~360}}=1.13.
\end{equation}

This value is between the contamination factor calculated from \citet{guzzo13} and the one given by our SVM opt+NIR classifier.
Taking this difference into account, we conclude that the results obtained by both methods are similar and very close to the real values obtained from the spectroscopic observations. 

We performed a classification of 264 objects preclassified as galaxies through  the SEDs+r$\rm{_h}$ method, and spectroscopicaly classified as a stars. 
In total 122 sources from this sample  (46.6\%) were correctly classified as stars by our algorithm. 
Taking  only sources with very high confidence level of spectroscopic classification into account ($\rm{VIPERS_{Zflag}}\geqslant$3; 123 sources),  we found that our algorithm shows 74.8\% of accuracy in correctly classifying 92 of those objects as stars.

It confirms that our SVM classification, based on spectroscopically measured objects from the VIPERS and VVDS surveys, can provide an efficient star/no-star classifier.
This method is also very fast.
The only time-consuming part of the SVM-based method is the tuning of the classifier, but once the classifier is trained, all the following classifications are very fast and can be done without any additional supervision.

\section{Conclusions}
\label{conclusions}
Application of the support vector machine algorithm can deliver an excellent (with accuracy level for self-check test higher that 98\% for galaxies, 94\% for AGNs, and 93\% for stars) classification  for three classes of objects, after a careful selection of the training sample. 
For our analysis we constructed two classifiers, with and without near infrared data using a multidimensional colour hyperspace. 
A part of the AGN and  star samples were extracted from the VVDS survey.
We have found a significant improvement in the SVM classification (8\% in the total accuracy of the classifier) adding an NIR  colour parameters to our feature vectors.

For the  optical+NIR classifier, we obtained very  good agreement (93.60\%, 81.80\%, and 92.52\% for galaxies, AGNs, and stars, respectively)  with the VIPERS spectroscopic sample with flag confidence level of z measurements equal to 95\%. 
What makes our approach to SVM classification more suitable is that  the enormous amount of excellent  quality data, means that we could create the classifier, which was trained on the part of the most secure sources, and then test it against the remaining secure objects to create the most efficient pattern recognition system. 
The VIPERS survey gathered a  large number of sources (55~358) with very good spectroscopic measurements, which then were strictly analysed to obtain the most secure redshifts.
This allowed for the choice  of the best sample,  which could be used as a basis  for the new methods of automatic classification. 

SVM classifiers are mostly used in the literature for separating of two classes of sources (e.g. stars and galaxies). 
The only recent application of the SVM to the galaxy/AGNs/stars classification was performed by \citet{saglia12}, who trained and used his classifier for the Pan-STARRS1 data. 
Comparing the accuracies of our classifier and those of \citet{saglia12} we found that our self-check results look somewhat better (97\%, 95\%, 97\% vs 97\%, 84\%, 85\% for galaxies, AGNs, stars for VIPERS and Pan-STARSS1 classifier, respectively). 
However, we have to stress that both methods cannot be directly compared because of initial differences in both surveys. 
Pan-STARRS1 is a magnitude-limited survey, which implies a much higher variety of properties of all the sources it contains. 
In contrast, VIPERS was preselected to contain only 0.5$<$z$<$1.2 galaxies, witch assures that they form a much more distinct and better separated group in a multicolour space. 
This may facilitate a separation between galaxies and AGNs, as well as a part of stars that were re-introduced to the VIPERS target sample as AGN candidates. 
On the other hand, the lack of `typical' stars in the VIPERS database (rejected after colour and half-light radius preselection) occupying the same colour-colour space as galaxies may hamper our classification based only on colours, and decrease the efficiency of our classifier for sources from the real sample.   
The difference in the performance with respect to the PAN-STARS1 SVM method might also be related to the different broad-band photometry. 
The tests of accuracy of our purely optical (u$^*$g'r'i') classifier show similar efficiency to the  PAN-STARS1 results (94\%, 82\%, and 93\% for galaxies, AGNs, and stars from VIPERS survey), while the dimension of PAN-STARS1 parameter space is higher than ours (4D in case of PAN-STARS1 and 3D in the case of VIPERS optical classifier). 
It suggests that the key points of our method might be a more suitable photometry (u* instead of z$\rm{_{P1}}$ and y$\rm{_{P1}}$ bands) and division of our sample into apparent magnitude bins.

Our  approach allows us to photometrically classify sources in the VIPERS survey, augmenting the spectral information.  
By classifying the sources with low-quality spectra, we can improve the classification and  enlarge the samples that may be used for analysis.  
Using the optical+NIR classifier, we confirmed the class of 4~900 objects with low flags. 
Further improvement in our classifier by the addition of the morphology and emission/absorption line information will improve the already very good performance of galaxy/AGN/star classifier.
It will also allow for developing of a more specific  galaxy and AGN-type classifications.

\begin{acknowledgements}
We acknowledge the crucial contribution of the ESO staff for the management of service observations. 
In particular, we are deeply grateful to M. Hilker for his constant help and support of this programme. 
Italian participation in VIPERS has been funded by INAF through the PRIN 2008 and 2010 programmes. 
LG and BRG acknowledge support by the European Research Council through the Darklight ERC Advanced Research Grant (\# 291521). 
OLF acknowledges the support of the European Research Council through the EARLY ERC Advanced Research Grant (\# 268107). 
Polish participants have been supported by the Polish Ministry of Science (grant N N203 51 29 38), the Polish-Swiss Astro Project (co-financed by a grant from Switzerland, through the Swiss Contribution to the enlarged European Union), the European Associated Laboratory Astrophysics Poland-France HECOLS, and the Japan Society for the Promotion of Science (JSPS) Postdoctoral Fellowship for Foreign Researchers (KM, P11802). 
AS has been supported by the Global COE Program Request for Fundamental Principles in the Universe: from Particles to the Solar System and the Cosmos commissioned by the Ministry of Education, Culture, Sports, Science and Technology (MEXT) of Japan, and by the JSPS Strategic Young Researcher Overseas Visits Program for Accelerating Brain Circulation, “Construction of a Global Platform for the Study of Sustainable Humanosphere. 
GDL acknowledges financial support from the European Research Council under the European Community's Seventh Framework Programme (FP7/2007-2013)/ERC grant agreement n. 202781. 
WJP and RT acknowledge financial support from the European Research Council under the European Community's Seventh Framework Programme (FP7/2007-2013)/ERC grant agreement n. 202686. 
WJP is also grateful for support from the UK Science and Technology Facilities Council through the grant ST/I001204/1. 
EB, FM, and LM acknowledge the support from grants ASI-INAF I/023/12/0 and PRIN MIUR 2010-2011. 
YM acknowledges support from CNRS/INSU (Institut National des Sciences de l’Univers) and the Programme National Galaxies et Cosmologie (PNCG). 
CM is grateful for support from specific project funding by the {\it Institut Universitaire de France} and the LABEX OCEVU.
\end{acknowledgements}

\bibliographystyle{aa} 


\begin{thebibliography}{60}
\expandafter\ifx\csname natexlab\endcsname\relax\def\natexlab#1{#1}\fi

\bibitem[{Akbani {et~al.}(2004)Akbani, Kwek, \& Japkowicz}]{akabani04}
Akbani, R., Kwek, S., \& Japkowicz, N. 2004, in In Proceedings of the 15th
  European Conference on Machine Learning (ECML), 39--50

\bibitem[{{Arnouts} {et~al.}(1999){Arnouts}, {Cristiani}, {Moscardini},
  {et~al.}}]{arnouts99}
{Arnouts}, S., {Cristiani}, S., {Moscardini}, L., {et~al.} 1999, MNRAS, 310,
  540

\bibitem[{{Ball} {et~al.}(2006){Ball}, {Brunner}, {Myers}, \&
  {Tcheng}}]{ball06}
{Ball}, N.~M., {Brunner}, R.~J., {Myers}, A.~D., \& {Tcheng}, D. 2006, \apj,
  650, 497

\bibitem[{{Beckwith} {et~al.}(2006){Beckwith}, {Stiavelli}, {Koekemoer},
  {Caldwell}, {Ferguson}, {Hook}, {Lucas}, {Bergeron}, {Corbin}, {Jogee},
  {Panagia}, {Robberto}, {Royle}, {Somerville}, \& {Sosey}}]{beckwith06}
{Beckwith}, S.~V.~W., {Stiavelli}, M., {Koekemoer}, A.~M., {et~al.} 2006, \aj,
  132, 1729

\bibitem[{Bel {et~al.}(2013)}]{bel13}
Bel, J. {et~al.} 2013, A\&A, , submitted

\bibitem[{{Bland-Hawthorn}(2012)}]{bland12}
{Bland-Hawthorn}, J. 2012, Research in Astronomy and Astrophysics, 12

\bibitem[{{Boulade} {et~al.}(2000){Boulade}, {Charlot}, {Abbon}, {Aune},
  {Borgeaud}, {Carton}, {Carty}, {Desforge}, {Eppele}, {Gallais}, {Gosset},
  {Granelli}, {Gros}, {de Kat}, {Loiseau}, {Mellier}, {Ritou}, {Rousse},
  {Starzynski}, {Vignal}, \& {Vigroux}}]{boulade00}
{Boulade}, O., {Charlot}, X., {Abbon}, P., {et~al.} 2000, in Society of
  Photo-Optical Instrumentation Engineers (SPIE) Conference Series, Vol. 4008,
  Society of Photo-Optical Instrumentation Engineers (SPIE) Conference Series,
  ed. M.~{Iye} \& A.~F. {Moorwood}, 657--668

\bibitem[{{Brightman} \& {Nandra}(2012)}]{brightman12}
{Brightman}, M. \& {Nandra}, K. 2012, \mnras, 422, 1166

\bibitem[{Chang \& Lin(2011)}]{chang11}
Chang, C.-C. \& Lin, C.-J. 2011, ACM Transactions on Intelligent Systems and
  Technology, 2, 27:1, software available at
  {http://www.csie.ntu.edu.tw/$\sim$cjlin/libsvm}, issue = {3}

\bibitem[{{Chiu} {et~al.}(2005){Chiu}, {Zheng}, {Schneider}, {Glazebrook},
  {Iye}, {Kashikawa}, {Tsvetanov}, {Yoshida}, \& {Brinkmann}}]{chiu05}
{Chiu}, K., {Zheng}, W., {Schneider}, D.~P., {et~al.} 2005, \aj, 130, 13


\bibitem[{{Colless} {et~al.}(2001){Colless},  {Dalton}, {Maddox}, {Sutherland}, {Norberg},  {Cole},  {Bland-Hawthorn},  {Bridges}, {Cannon}, {Collins},  {Couch},  {Cross}, {Deeley},  {De Propris}, {Driver}, {Efstathiou}, {Ellis}, {Frenk}, {Glazebrook},  {Jackson}, {Lahav},  {Lewis},  {Lumsden},  {Madgwick}, {Peacock},  {Peterson},  {Price},  {Seaborne} \&	{Taylor}}]{colless01} 
{Colless}, M., {Dalton}, G., {Maddox}, S., {et~al.} 2001, \mnras, 328, 1039

\bibitem[{{Colless} {et~al.}(2003){Colless}, {Peterson}, {Jackson}, {Peacock}, {Cole}, {Norberg}, {Baldry}, {Baugh}, {Bland-Hawthorn},  {Bridges},  {Cannon}, {Collins}, {Couch},  {Cross},  {Dalton}, {De~Propris},  {Driver},  {Efstathiou}, {Ellis},  {Frenk},  {Glazebrook},  {Lahav}, {Lewis}, {Lumsden},  {Maddox}, {Madgwick}, {Sutherland} \& {Taylor}}]{colless03}
{Colless}, M., {Peterson}, B.~A., {Jackson}, C., {et~al.} 2003, ArXiv:astro-ph/0306581

\bibitem[{{Coupon} {et~al.}(2009){Coupon}, {Ilbert}, {Kilbinger}, {McCracken},
  {Mellier}, {Arnouts}, {Bertin}, {Hudelot}, {Schultheis}, {Le F{\`e}vre}, {Le
  Brun}, {Guzzo}, {Bardelli}, {Zucca}, {Bolzonella}, {Garilli}, {Zamorani},
  {Zanichelli}, {Tresse} \& {Aussel}}]{coupon09}
{Coupon}, J., {Ilbert}, O., {Kilbinger}, M., {et~al.} 2009, \aap, 500, 981

\bibitem[{Cristianini \& Shawe-Taylor(2000)}]{crisrianini00}
Cristianini, N. \& Shawe-Taylor, J. 2000, An Introduction to Support Vector
  Machines: And Other Kernel-Based Learning Methods (Cambridge University
  Press)

\bibitem[{Davidzon {et~al.}(2013)Davidzon, {Bolzonella}, {et~al.}}]{davidzon13}
Davidzon, I., {Bolzonella}, M., {et~al.} 2013, A\&A, submitted,  ArXiv e-print 1303.3808

\bibitem[{{de la Torre} {et~al.}(2013){de la Torre}, {Guzzo}, {Peacock},
  Branchini, {Iovino}, \& {Granett}}]{delatorre13}
{de la Torre}, S., {Guzzo}, L., {Peacock}, J.~A., {et~al.} 2013, A\&A,
  submitted, ArXiv e-print 1303.2622
  
\bibitem[{Drinkwater {et~al.}(2003)Drinkwater, Gregg, Hilker, Bekki, Couch, Ferguson, Jones, \& Phillipps}]{drinkwater03}  
 {Drinkwater}, M.~J., {Gregg}, M.~D., {Hilker}, M., {et~al.} 2003, Nature 423, 519 
  
\bibitem[{{Emerson} \& {Sutherland}(2010)}]{emerson10}
{Emerson}, J. \& {Sutherland}, W. 2010, The Messenger, 139, 2

\bibitem[{{Fadely} {et~al.}(2012){Fadely}, {Hogg}, \& {Willman}}]{fadely12}
{Fadely}, R., {Hogg}, D.~W., \& {Willman}, B. 2012, \apj, 760, 15

\bibitem[{Fritz {et~al.}(2013)Fritz, {Scodeggio}, {et~al.}}]{fritz13}
Fritz, A., {Scodeggio}, M., {et~al.} 2013, A\&A, submitted

\bibitem[{{Garilli} {et~al.}(2008){Garilli}, {Le F{\`e}vre}, {Guzzo},
  {Maccagni}, {Le Brun}, {de la Torre}, {Meneux}, {Tresse}, {Franzetti},
  {Zamorani}, {Zanichelli}, {Gregorini}, {Vergani}, {Bottini}, {Scaramella},
  {Scodeggio}, {Vettolani}, {Adami}, {Arnouts}, {Bardelli}, {Bolzonella},
  {Cappi}, {Charlot}, {Ciliegi}, {Contini}, {Foucaud}, {Gavignaud}, {Ilbert},
  {Iovino}, {Lamareille}, {McCracken}, {Marano}, {Marinoni}, {Mazure},
  {Merighi}, {Paltani}, {Pell{\`o}}, {Pollo}, {Pozzetti}, {Radovich}, {Zucca},
  {Blaizot}, {Bongiorno}, {Cucciati}, {Mellier}, {Moreau}, \&
  {Paioro}}]{garilli08}
{Garilli}, B., {Le F{\`e}vre}, O., {Guzzo}, L., {et~al.} 2008, \aap, 486, 683

\bibitem[{{Garilli} {et~al.}(2012){Garilli}, {Paioro}, {Scodeggio},
  {Franzetti}, {Fumana}, \& {Guzzo}}]{garilli12}
{Garilli}, B., {Paioro}, L., {Scodeggio}, M., {et~al.} 2012, \pasp, 124, 1232

\bibitem[{{Gavignaud} {et~al.}(2007){Gavignaud}, {Bongiorno}, {Paltani},
  {Mathez}, {Zamorani}, {Moller}, {Picat}, {Le Brun}, {Marano}, {Le Fevre},
  {Bottini}, {Garilli}, {Maccagni}, {Scaramella}, {Scodeggio}, {Tresse},
  {Vettolani}, {Zanichelli}, {Adami}, {Arnaboldi}, {Arnouts}, {Bardelli},
  {Bolzonella}, {Cappi}, {Charlot}, {Ciliegi}, {Contini}, {Foucaud},
  {Franzetti}, {Guzzo}, {Ilbert}, {Iovino}, {McCracken}, {Marinoni}, {Mazure},
  {Meneux}, {Merighi}, {Pello}, {Pollo}, {Pozzetti}, {Radovich}, {Zucca},
  {Bondi}, {Busarello}, {Cucciati}, {de La Torre}, {Gregorini}, {Lamareille},
  {Mellier}, {Merluzzi}, {Ripepi}, {Rizzo}, \& {Vergani}}]{gavignaud07}
{Gavignaud}, I., {Bongiorno}, A., {Paltani}, S., {et~al.} 2007, VizieR Online
  Data Catalog, 345, 70079

\bibitem[{{Goranova} {et~al.}(2009){Goranova}, {Hudelot}, {Contini}, {Magnard},
  McCracken, Mellier, Monnerville, Schultheis, Semah, Cuillandre, \&
  Aussel}]{goranova09}
{Goranova}, Y., {Hudelot}, P., {Contini}, T., {et~al.} 2009, The CFHTLS T0006
  Release, http://terapix.iap.fr/cplt/table\_syn\_T0006.html

\bibitem[{Guzzo {et~al.}(2013)Guzzo, {Scodeggio}, {Garilli}, {Granett},
  {Abbas}, {et~al.}}]{guzzo13}
Guzzo, L., {Scodeggio}, M., {Garilli}, B., {et~al.} 2013, A\&A, submitted, ArXiv e-print 1303.2623

\bibitem[{{Hassan} {et~al.}(2013){Hassan}, {Mirabal}, {Contreras}, \&
  {Oya}}]{hassan13}
{Hassan}, T., {Mirabal}, N., {Contreras}, J.~L., \& {Oya}, I. 2013, \mnras,
  428, 220

\bibitem[{{Henrion} {et~al.}(2011){Henrion}, {Mortlock}, {Hand}, \&
  {Gandy}}]{henrion11}
{Henrion}, M., {Mortlock}, D.~J., {Hand}, D.~J., \& {Gandy}, A. 2011, \mnras,
  412, 2286

\bibitem[{Hsu {et~al.}(2010)Hsu, Chang, \& C.-J.}]{hsu10}
Hsu, C.-W., Chang, C.~C., \& C.-J., L. 2010, A Practical Guide to Support
  Vector Classification, Department of Computer Science, National Taiwan
  University, Taiwan, http://www.csie.ntu.edu.tw/~cjlin/papers/guide/guide.pdf

\bibitem[{{Huertas-Company} {et~al.}(2008){Huertas-Company}, {Rouan}, {Tasca},
  {Soucail}, \& {Le F{\`e}vre}}]{huertas08}
{Huertas-Company}, M., {Rouan}, D., {Tasca}, L., {Soucail}, G., \& {Le
  F{\`e}vre}, O. 2008, \aap, 478, 971

\bibitem[{{Ilbert} {et~al.}(2006){Ilbert}, {Arnouts}, {McCracken},
  {et~al.}}]{ilbert06}
{Ilbert}, O., {Arnouts}, S., {McCracken}, H.~J., {et~al.} 2006, A\&A, 457, 841

\bibitem[{{Ivezic} {et~al.}(2009){Ivezic}, {Tyson}, {Axelrod}, {Burke},
  {Claver}, {Cook}, {Kahn}, {Lupton}, {Monet}, {Pinto}, {Strauss}, {Stubbs},
  {Jones}, {Saha}, {Scranton}, {Smith}, \& {LSST Collaboration}}]{ivezic09}
{Ivezic}, Z., {Tyson}, J.~A., {Axelrod}, T., {et~al.} 2009, in Bulletin of the
  American Astronomical Society, Vol.~41, American Astronomical Society Meeting
  Abstracts 213, 460.03

\bibitem[{{Kaiser} {et~al.}(2010){Kaiser}, {Burgett}, {Chambers}, {Denneau},
  {Heasley}, {Jedicke}, {Magnier}, {Morgan}, {Onaka}, \& {Tonry}}]{kaiser10}
{Kaiser}, N., {Burgett}, W., {Chambers}, K., {et~al.} 2010, in Society of
  Photo-Optical Instrumentation Engineers (SPIE) Conference Series, Vol. 7733,
  Society of Photo-Optical Instrumentation Engineers (SPIE) Conference Series

\bibitem[{{Kron}(1980)}]{kron80}
{Kron}, R.~G. 1980, \apjs, 43, 305

\bibitem[{{Laureijs} {et~al.}(2012){Laureijs}, {Gondoin}, {Duvet}, {Saavedra
  Criado}, {Hoar}, {Amiaux}, {Augu{\`e}res}, {Cole}, {Cropper}, {Ealet},
  {Ferruit}, {Escudero Sanz}, {Jahnke}, {Kohley}, {Maciaszek}, {Mellier},
  {Oosterbroek}, {Pasian}, {Sauvage}, {Scaramella}, {Sirianni}, \&
  {Valenziano}}]{laureijis12}
{Laureijs}, R., {Gondoin}, P., {Duvet}, L., {et~al.} 2012, in Society of
  Photo-Optical Instrumentation Engineers (SPIE) Conference Series, Vol. 8442,
  Society of Photo-Optical Instrumentation Engineers (SPIE) Conference Series

\bibitem[{{Le F{\`e}vre} {et~al.}(2000){Le F{\`e}vre}, {Saisse}, {Mancini},
  {Vettolani}, {Maccagni}, {Picat}, {Mellier}, {Mazure}, {Cuby}, {Delabre},
  {Garilli}, {Hill}, {Prieto}, {Voet}, {Arnold}, {Brau-Nogue}, {Cascone},
  {Conconi}, {Finger}, {Huster}, {Laloge}, {Lucuix}, {Mattaini}, {Schipani},
  {Waultier}, {Zerbi}, {Avila}, {Beletic}, {D'Odorico}, {Moorwood}, {Monnet},
  \& {Reyes Moreno}}]{lefevre00}
{Le F{\`e}vre}, O., {Saisse}, M., {Mancini}, D., {et~al.} 2000, in Society of
  Photo-Optical Instrumentation Engineers (SPIE) Conference Series, Vol. 4008,
  Society of Photo-Optical Instrumentation Engineers (SPIE) Conference Series,
  ed. M.~{Iye} \& A.~F. {Moorwood}, 546--557

\bibitem[{{Le F{\`e}vre} {et~al.}(2005){Le F{\`e}vre}, {Vettolani}, {Garilli},
  {Tresse}, {Bottini}, {Le Brun}, {Maccagni}, {Picat}, {Scaramella},
  {Scodeggio}, {Zanichelli}, {Adami}, {Arnaboldi}, {Arnouts}, {Bardelli},
  {Bolzonella}, {Cappi}, {Charlot}, {Ciliegi}, {Contini}, {Foucaud},
  {Franzetti}, {Gavignaud}, {Guzzo}, {Ilbert}, {Iovino}, {McCracken}, {Marano},
  {Marinoni}, {Mathez}, {Mazure}, {Meneux}, {Merighi}, {Paltani}, {Pell{\`o}},
  {Pollo}, {Pozzetti}, {Radovich}, {Zamorani}, {Zucca}, {Bondi}, {Bongiorno},
  {Busarello}, {Lamareille}, {Mellier}, {Merluzzi}, {Ripepi}, \&
  {Rizzo}}]{lefevre05}
{Le F{\`e}vre}, O., {Vettolani}, G., {Garilli}, B., {et~al.} 2005, \aap, 439,
  845

\bibitem[{{Le F{\`e}vre} {et~al.}(2013, in preparation)}]{lefevre13}
{Le F{\`e}vre}, O. {et~al.} 2013, in preparation, http://cesam.oamp.fr/vuds

\bibitem[{{Marchetti} {et~al.}(2012){Marchetti}, {Granett}, {Guzzo}, {Fritz},
  {Garilli}, {Scodeggio}, {Abbas}, {Adami}, {Arnouts}, {Bolzonella}, {Bottini},
  {Cappi}, {Coupon}, {Cucciati}, {De Lucia}, {de la Torre}, {Franzetti},
  {Fumana}, {Ilbert}, {Iovino}, {Krywult}, {Le Brun}, {Le Fevre}, {Maccagni},
  {Malek}, {Marulli}, {McCracken}, {Meneux}, {Paioro}, {Polletta}, {Pollo},
  {Schlagenhaufer}, {Tasca}, {Tojeiro}, {Vergani}, {Zanichelli}, {Bel},
  {Bersanelli}, {Blaizot}, {Branchini}, {Burden}, {Davidzon}, {Porto},
  {Guennou}, {Marinoni}, {Mellier}, {Moscardini}, {Nichol}, {Peacock},
  {Percival}, {Phleps}, {Schimd}, {Wolk}, \& {Zamorani}}]{marchetti12}
{Marchetti}, A., {Granett}, B.~R., {Guzzo}, L., {et~al.} 2012, \mnras, 107

\bibitem[{Marulli {et~al.}(2013)Marulli, {Bolzonella}, Branchini, Davidzon, {de
  la Torre}, {Guzzo}, {Iovino}, {Moscardini}, {Pollo}, {Abbas}, {Adami},
  {Arnouts}, Bel, Bottini, Cappi, {et~al.}}]{marulli13}
Marulli, F., {Bolzonella}, M., Branchini, E., {et~al.} 2013, A\&A, submitted,	ArXiv e-print 1303.2633

\bibitem[{Mellier {et~al.}(2008)Mellier, Bertin, Hudelot, {et~al.}}]{mellier08}
Mellier, Y., Bertin, E., Hudelot, P., {et~al.} 2008, The CFHTLS T0005 Release,
  http://terapix.iap.fr/cplt/oldSite/Descart/CFHTLS-T0005-Release.pdf

\bibitem[{Meyer(2001)}]{meyer01}
Meyer, D. 2001, R News, 1, 23

\bibitem[{Mohr {et~al.}(2012)Mohr, Armstrong, Bertin, {Daues}, {Desai},  {Gower},  {Gruendl},  {Hanlon}, {Kuropatkin},  {Lin}, {Marriner},  {Petravic}, {Sevilla},  {Swanson}, {Tomashek},  {Tucker} \& 
	{Yanny}}]{mohr12}
{Mohr}, J.~J. and {Armstrong}, R. and {Bertin}, E., {et~al.} 2012, in {Software and Cyberinfrastructure for Astronomy II. }, Vol.~8451, {Society of Photo-Optical Instrumentation Engineers (SPIE) Conference Series}, 

\bibitem[{{Peng} {et~al.}(2012){Peng}, {Zhang}, {Zhao}, \& {Wu}}]{peng12}
{Peng}, N., {Zhang}, Y., {Zhao}, Y., \& {Wu}, X.-b. 2012, \mnras, 425, 2599

\bibitem[{{Pierre} {et~al.}(2004){Pierre}, {Valtchanov}, {Altieri}, {Andreon},
  {Bolzonella}, {Bremer}, {Disseau}, {Dos Santos}, {Gandhi}, {Jean}, {Pacaud},
  {Read}, {Refregier}, {Willis}, {Adami}, {Alloin}, {Birkinshaw}, {Chiappetti},
  {Cohen}, {Detal}, {Duc}, {Gosset}, {Hjorth}, {Jones}, {Le F{\`e}vre},
  {Lonsdale}, {Maccagni}, {Mazure}, {McBreen}, {McCracken}, {Mellier},
  {Ponman}, {Quintana}, {Rottgering}, {Smette}, {Surdej}, {Starck}, {Vigroux},
  \& {White}}]{pierre04}
{Pierre}, M., {Valtchanov}, I., {Altieri}, B., {et~al.} 2004, \jcap, 9, 11

\bibitem[{{Pollo} {et~al.}(2010){Pollo}, {Rybka}, \& {Takeuchi}}]{pollo10}
{Pollo}, A., {Rybka}, P., \& {Takeuchi}, T.~T. 2010, \aap, 514, A3

\bibitem[{{Puget} {et~al.}(2004){Puget}, {Stadler}, {Doyon}, {Gigan},
  {Thibault}, {Luppino}, {Barrick}, {Benedict}, {Forveille}, {Rambold},
  {Thomas}, {Vermeulen}, {Ward}, {Beuzit}, {Feautrier}, {Magnard}, {Mella},
  {Preis}, {Vallee}, {Wang}, {Lin}, {Hall}, \& {Hodapp}}]{puget04}
{Puget}, P., {Stadler}, E., {Doyon}, R., {et~al.} 2004, in Society of
  Photo-Optical Instrumentation Engineers (SPIE) Conference Series, Vol. 5492,
  Society of Photo-Optical Instrumentation Engineers (SPIE) Conference Series,
  ed. A.~F.~M. {Moorwood} \& M.~{Iye}, 978--987

\bibitem[{Raskutti \& Kowalczyk(2004)}]{raskutti04}
Raskutti, B. \& Kowalczyk, A. 2004, SIGKDD Explor. Newsl., 6, 60

\bibitem[{{Richards} {et~al.}(2002){Richards}, {Fan}, {Newberg}, {Strauss},
  {Vanden Berk}, {Schneider}, {Yanny}, {Boucher}, {Burles}, {Frieman}, {Gunn},
  {Hall}, {Ivezi{\'c}}, {Kent}, {Loveday}, {Lupton}, {Rockosi}, {Schlegel},
  {Stoughton}, {SubbaRao}, \& {York}}]{richards02}
{Richards}, G.~T., {Fan}, X., {Newberg}, H.~J., {et~al.} 2002, \aj, 123, 2945

\bibitem[{{Saglia} {et~al.}(2012){Saglia}, {Tonry}, {Bender}, {Greisel},  {Seitz}, {Senger}, {Snigula}, {Phleps}, {Wilman}, {Bailer-Jones}, {Klement},  {Rix}, {Smith}, {Green}, {Burgett}, {Chambers}, {Heasley}, {Kaiser}, {Magnier}, {Morgan}, {Price}, {Stubbs}, \& {Wainscoat}}]{saglia12}
  {Saglia}, R.~P., {Tonry}, J.~L., {Bender}, R., {et~al.} 2012, \apj, 746, 128

\bibitem[{Schlegel} {et.~al.}(1998){Schlegel}, {Finkbeiner} \& {Davis}]{schlegel98}
{Schlegel}, D.~J., {Finkbeiner}, D.~P., {Davis}, M., 1998 \apj, 500, 525


\bibitem[{{Scodeggio} {et~al.}(2009){Scodeggio}, {Franzetti}, {Garilli}, {Le
  F{\`e}vre}, \& {Guzzo}}]{scodeggio09}
{Scodeggio}, M., {Franzetti}, P., {Garilli}, B., {Le F{\`e}vre}, O., \&
  {Guzzo}, L. 2009, The Messenger, 135, 13

\bibitem[{Shawe-Taylor \& Cristianini(2004)}]{shawe04}
Shawe-Taylor, J. \& Cristianini, N. 2004, Kernel Methods for Pattern Analysis
  (Cambridge University Press)

  \bibitem[{{Sholl} {et~al.}(2012)Sholl, Ackerman,  Bebek, Besuner, Dey, Edelstein,  Jelinsky,  Lampton,  Levi, Liang,  Perry,  Roe,  Silber,  Schlegel}]{sholl12}
Sholl M.~J., Ackerman~M.~R., Bebek~C.,  {et~al.} 2012, in Ground-based and Airborne Instrumentation for Astronomy IV, Proceedings of the SPIE, Vol. 8446, 844667

  
\bibitem[{{Solarz} {et~al.}(2012){Solarz}, {Pollo}, {Takeuchi}, {P{\c e}piak},
  {Matsuhara}, {Wada}, {Oyabu}, {Takagi}, {Goto}, {Ohyama}, {Pearson},
  {Hanami}, \& {Ishigaki}}]{solarz12}
{Solarz}, A., {Pollo}, A., {Takeuchi}, T.~T., {et~al.} 2012, \aap, 541, A50

\bibitem[{{Stern} {et~al.}(2012){Stern}, {Assef}, {Benford}, {Blain}, {Cutri},
  {Dey}, {Eisenhardt}, {Griffith}, {Jarrett}, {Lake}, {Masci}, {Petty},
  {Stanford}, {Tsai}, {Wright}, {Yan}, {Harrison}, \& {Madsen}}]{stern12}
{Stern}, D., {Assef}, R.~J., {Benford}, D.~J., {et~al.} 2012, \apj, 753, 30

\bibitem[{{Stern} {et~al.}(2005){Stern}, {Eisenhardt}, {Gorjian}, {Kochanek},
  {Caldwell}, {Eisenstein}, {Brodwin}, {Brown}, {Cool}, {Dey}, {Green},
  {Jannuzi}, {Murray}, {Pahre}, \& {Willner}}]{stern05}
{Stern}, D., {Eisenhardt}, P., {Gorjian}, V., {et~al.} 2005, \apj, 631, 163

\bibitem[{Tang {et~al.}(2009)Tang, Zhang, Chawla, \& Krasser}]{tang09}
Tang, Y., Zhang, Y.-Q., Chawla, N.~V., \& Krasser, S. 2009, IEEE Transactions
  on Systems, Man, and Cybernetics, Part B (Cybernetics), 39, 281

\bibitem[{{Thibault} {et~al.}(2003){Thibault}, {Cui}, {Poirier}, {Vallee},
  {Doyon}, {Rabou}, \& {Salmon}}]{thibault03}
{Thibault}, S., {Cui}, Q., {Poirier}, M., {et~al.} 2003, in Society of
  Photo-Optical Instrumentation Engineers (SPIE) Conference Series, Vol. 4841,
  Society of Photo-Optical Instrumentation Engineers (SPIE) Conference Series,
  ed. M.~{Iye} \& A.~F.~M. {Moorwood}, 932--943

\bibitem[{Tian {et~al.}(2011)Tian, Gu, \& Liu}]{tian11}
Tian, J., Gu, H., \& Liu, W. 2011, Neural Comput. Appl., 20, 203

\bibitem[{Vapnik(1995)}]{vapnik95}
Vapnik, V.~N. 1995, The Nature of Statistical Learning Theory (Springer)


\bibitem[Vanschoenwinkel \&  Manderick (2005)]{vanschoenwinkel05}
Vanschoenwinkel B. \&  Manderick B., 2005, in Proceedings of the First international conference on Deterministic and Statistical Methods in Machine Learning, 256

\bibitem[{{Vasconcellos} {et~al.}(2011){Vasconcellos}, {de Carvalho}, {Gal},  {LaBarbera}, {Capelato}, {Frago Campos Velho}, {Trevisan}, \&  {Ruiz}}]{vasconcellos11}
{Vasconcellos}, E.~C., {de Carvalho}, R.~R., {Gal}, R.~R., {et~al.} 2011, \aj,  141, 189
  

\bibitem[{{Walker} {et~al.}(1989){Walker}, {Volk}, {Wainscoat}, {Schwartz}, \&
  {Cohen}}]{walker89}
{Walker}, H.~J., {Volk}, K., {Wainscoat}, R.~J., {Schwartz}, D.~E., \& {Cohen},
  M. 1989, \aj, 98, 2163

\bibitem[{{Wittman} {et~al.}(2002){Wittman}, {Tyson}, {Dell'Antonio}, {Becker},
  {Margoniner}, {Cohen}, {Norman}, {Loomba}, {Squires}, {Wilson}, {Stubbs},
  {Hennawi}, {Spergel}, {Boeshaar}, {Clocchiatti}, {Hamuy}, {Bernstein},
  {Gonzalez}, {Guhathakurta}, {Hu}, {Seljak}, \& {Zaritsky}}]{wittman02}
{Wittman}, D.~M., {Tyson}, J.~A., {Dell'Antonio}, I.~P., {et~al.} 2002, in
  Society of Photo-Optical Instrumentation Engineers (SPIE) Conference Series,
  Vol. 4836, Society of Photo-Optical Instrumentation Engineers (SPIE)
  Conference Series, ed. J.~A. {Tyson} \& S.~{Wolff}, 73--82

\bibitem[{{Wo{\'z}niak} {et~al.}(2004){Wo{\'z}niak}, {Williams}, {Vestrand}, \&
  {Gupta}}]{wozniak04}
{Wo{\'z}niak}, P.~R., {Williams}, S.~J., {Vestrand}, W.~T., \& {Gupta}, V.
  2004, \aj, 128, 2965

\end{thebibliography}

\end{document}